\documentclass[11pt,a4paper]{article}
\pdfoutput=1
\usepackage{jheppub}
\usepackage[T1]{fontenc}
\usepackage{slashed}
\usepackage{color}
\usepackage{graphicx}
\usepackage{amsmath}
\usepackage{amsfonts}
\usepackage{amssymb}
\usepackage{amsmath}
\usepackage{amssymb}
\usepackage{wasysym}
\usepackage{bm}
\usepackage{latexsym}
\usepackage{stmaryrd}
\usepackage{textcomp}
\usepackage{verbatim}
\usepackage{slashed}
\usepackage{array}

\usepackage[normalem]{ulem}
\usepackage{acronym,epsfig,subfigure,skak}
\usepackage{latexsym,amsfonts,amssymb,amsmath,makeidx,mathrsfs,multirow,
lineno,bm,bbm,fancyvrb,ifpdf}

\allowdisplaybreaks

\newcommand{\blue}{\textcolor{blue}}

\catcode`@=12

\newcommand{\lsim}{\buildrel < \over {_\sim}}
\newcommand{\gsim}{\buildrel > \over {_\sim}}
\newcommand{\be}{\begin{equation}}
\newcommand{\ee}{\end{equation}}
\newcommand{\bea}{\begin{eqnarray}}
\newcommand{\eea}{\end{eqnarray}}
\newcommand{\ba}{\begin{array}}
\newcommand{\ea}{\end{array}}

\def\lsim{\mathrel{\raise.3ex\hbox{$<$\kern-.75em\lower1ex\hbox{$\sim$}}}}
\def\gsim{\mathrel{\raise.3ex\hbox{$>$\kern-.75em\lower1ex\hbox{$\sim$}}}}

\def\gev{\,{\rm GeV}}

\def\to{\rightarrow}

\def\beq{\begin{equation}}
\def\eeq{\end{equation}}
\def\be{\begin{equation}}
\def\ee{\end{equation}}
\def\bea{\begin{eqnarray}}
\def\eea{\end{eqnarray}}
\def\gev{\,{\rm GeV}}

\def\to{\rightarrow}

\def\beq{\begin{equation}}
\def\eeq{\end{equation}}

\begin{document}
\title{Gravitational waves triggered by $B-L$ charged hidden scalar and leptogenesis }

\author[a]{~Ligong Bian,}
\emailAdd{lgbycl@cqu.edu.cn}
\author[b]{Wei Cheng,}
\emailAdd{chengwei@itp.ac.cn}
\author[c]{Huai-Ke Guo,}
\emailAdd{ghk@ou.edu}
\author[d,e]{Yongchao Zhang}
\emailAdd{yongchao.zhang@physics.wustl.edu}

\affiliation[a]{Department of Physics, Chongqing University, Chongqing 401331, China}
\affiliation[b]{State Key Laboratory of Theoretical Physics, Institute of Theoretical Physics,
Chinese Academy of Sciences, Beijing, 100190, China}
\affiliation[c]{Department of Physics and Astronomy, University of Oklahoma, Norman, OK 73019, USA}
\affiliation[d]{Department of Physics and McDonnell Center for the Space Sciences, Washington University, St.Louis, MO 63130, USA}
\affiliation[e]{Center for High Energy Physics, Peking University, Beijing 100871, China}

\date{\today}

%\begin{abstract}
\abstract{
%We study the electroweak symmetry breaking from a dynamical breaking of a classically conformal $U(1)_{B-L}$ theory. The $B-L$ charged hidden scalar and right handed neutrinos are introduced to realize the phase transition. The gravitational wave signals generated are found to be probable by the future space-based interferometer LISA. The baryon asymmetry of the Universe is addressed by the leptogenesis where the dilution process induced by the hidden scalar is investigated.
%\end{abstract}
We study the electroweak symmetry breaking in the framework of a classically conformal $U(1)_{B-L}$ theory, where three right-handed neutrinos (RHNs) and a hidden scalar are introduced, with the latter playing the role of dark matter (DM). It is found that the DM and RHN sectors are crucial for the spontaneous symmetry breaking of the $U(1)_{B-L}$ symmetry, strong first order phase transition in the conformal theory and the resultant gravitational wave (GW) prospects at future space-based interferometer LISA and other GW experiments. The baryon asymmetry of the Universe is addressed by the resonant leptogenesis mechanism, which is potentially disturbed by the hidden scalar. To make the GW spectra detectable by LISA and resonant leptogenesis work in the conformal $U(1)_{B-L}$ theory, the hidden scalar can not fully saturate the observed DM relic density.}
%For the supercool dark matter situation scenario, we refer to Ref.~\cite{Hambye:2018qjv}. Where, the phase transition and leptogenesis mechanism are also different.

\maketitle

\section{Introduction}

Since the discovery of the standard model (SM) Higgs at the Large Hadron Collider (LHC), understanding the hierarchy problem becomes one of the most challenging theoretical difficulties in the SM, i.e. why the SM Higgs mass is much lower than the Planck scale? In light of null result in searches of new heavy particles at LHC, in particular the supersymmetric particles, the hierarchy problem is getting more concerned. This is also intimately related to the spontaneous electroweak symmetry breaking (EWSB), which is responsible for the generation of SM particle masses, and underlying more fundamental theories. One elegant way out is the Coleman-Weinberg mechanism~\cite{Coleman:1973jx}, in which the original potential is classically conformal and the EWSB is induced when the mass term is generated radiatively. As the conformal scale invariant version of the SM is not consistent with the Higgs data, one may take advantage of the Higgs-portal and therefore obtain a natural EWSB~\cite{Englert:2013gz,Farzinnia:2013pga}. The possibility to accommodate dark matter (DM) particles and inflation has been considered~\cite{Khoze:2013uia}, where extra scalar fields are introduced which are charged under the conformal $U(1)_{B-L}$ gauge group extension of the SM and are viable DM candidates~\cite{Iso:2009ss}.
% i.e. the classically conformal $B-L$ extended SM

The $U(1)_{B-L}$ symmetry breaking process could be dynamical while the Universe cools down, i.e. being a phase transition process. When the phase transition is first order, gravitational waves (GWs) could be generated and detected in current and future GW experiments,  such as LISA~\cite{Audley:2017drz, Cornish:2018dyw}, Taiji~\cite{Gong:2014mca}, TianQin~\cite{Luo:2015ght}, Big Bang Observer (BBO)~\cite{Corbin:2005ny}, DECi-hertz Interferometer Gravitational wave Observatory (DECIGO)~\cite{Musha:2017usi} and Ultimate-DECIGO~\cite{Kudoh:2005as}. For the study of the GW signal predictions within this framework, see e.g.,
Ref.~\cite{Ellis:2019oqb,Ellis:2018mja,Jinno:2016knw,Iso:2017uuu,Chao:2017vrq,Chao:2017ilw}. Where the $B-L$ symmetry can break at the TeV scale or under QCD scale after the QCD phase transition.

For the purpose of gauge anomaly cancellation, three right-handed neutrinos (RHNs) $N_i$ (with $i = 1,\,2,\,3$) are introduced to the $U(1)_{B-L}$ model, which can be used to generate the tiny neutrino masses via the type-I seesaw mechanism~\cite{Minkowski:1977sc, Mohapatra:1979ia, Yanagida:1979as, GellMann:1980vs, Glashow:1979nm}. Furthermore, the lepton asymmetry can be generated from the CP violating decays of heavy RHNs, i.e. through the mechanism of leptongenesis~\cite{Fukugita:1986hr}, which is then transferred into the baryon asymmetry through electroweak sphaleron processes. For the studies of leptogenesis in conformal theories, see e.g. Ref.~\cite{Khoze:2013oga}. If only one RHN is involved in leptogenesis, the RHN is too heavy to be produced at colliders, and we consider the TeV-scale resonant leptogenesis with two mass quasi-degenerate RHNs~\cite{Covi:1996wh, Flanz:1996fb, Pilaftsis:1997jf, Pilaftsis:2003gt}. As the RHNs couples to the scalar $\Phi$ and the heavy $Z'$ gauge boson which is from the $U(1)_{B-L}$ symmetry breaking, the processes $NN \to f\bar{f},\, \Phi\Phi,\, \Phi Z'$ (with $f$ the SM fermions) will dilute the heavy RHNs by two units, thus reducing the lepton and baryon asymmetry significantly~\cite{Blanchet:2009bu, Blanchet:2010kw, Iso:2010mv, Okada:2012fs, Heeck:2016oda, Dev:2017xry}.
%conformal scalar $\Phi$, whose quantum number is $2$ under the $U(1)_{B-L}$ gauge group (see Table~\ref{tab:content}). When $\Phi$ gets a non-vanishing vacuum expectation value (VEV), it will break the $U(1)_{B-L}$ gauge symmetry, and GW can be emitted if the phase transition is first order, including the three sources of collisions of the bubbles~\cite{Kosowsky:1991ua, Kosowsky:1992rz, Kosowsky:1992vn, Huber:2008hg, Jinno:2016vai, Jinno:2017fby}, bulk motion of the plasma in the form of sound waves (SWs)~\cite{Hindmarsh:2013xza,Hindmarsh:2015qta} and Magnetohydrodynamic (MHD) turbulence~\cite{Caprini:2009yp, Binetruy:2012ze} (see Refs.~\cite{Caprini:2015zlo, Cai:2017cbj, Weir:2017wfa} for recent reviews). \blue{Since we work in the classical invariant case, we can expect the symmetry breaking is driven by the gauge coupling $g_{BL}$ or the $U(1)_{B-L}$, here it is the SM-DM coupling $\lambda_{s\phi}$.
%including gravitational wave emission, dark matter, neutrino mass generation via the type-I seesaw and baryon asymmetry from leptogenesis

When a scalar ${\cal S}$ with nontrivial $B-L$ charge is introduced to the $U(1)_{B-L}$ model, it could be a viable DM candidate if it does not develop a non-vanishing vacuum expectation value (VEV). See Ref.~\cite{Hambye:2018qjv} for a supercool DM scenario where the GW signal from phase transition is quite different from our scenario. In this work, the scalar ${\cal S}$ has  nontrivial $B-L$ charges and its $B-L$ charge should be $n_{x}\neq \pm2 n$ with $n$ being an integer and smaller than $4$ such that the neutral DM scalar can be stabilized by the accidental $B-L$ symmetry~\cite{Rodejohann:2015lca}. As a result of the $B-L$ charge of ${\cal S}$,  the leptogenesis diffusion process can be disturbed and the annihilation process $NN \to {\cal S} {\cal S}^\dagger$ is also important. This dilution effect falsifies leptogenesis in a large region of parameter space (see Fig.~\ref{fig:leptogenesis}).  Even though the DM scalar ${\cal S}$ and and its complex conjugate ${\cal S}^\dagger$ can be pair produced through both the scalar and gauge portals, the monojet and other DM searches at the LHC are too weak to exclude any parameter space~\cite{Khachatryan:2014rra, Aad:2015zva, Aaboud:2017phn, Sirunyan:2017jix}. However, we find that the DM scalar ${\cal S}$ and the RHNs and their couplings play an important role in the phase transition and GW emission. We estimate the possibility of whether the DM (hidden) scalar can saturate the DM relic abundance and at the same time satisfy the current limits from low-background direct DM searches, i.e. those from LUX~\cite{Akerib:2016vxi}, PandaX-II~\cite{Tan:2016zwf, Cui:2017nnn} and Xenon1T~\cite{Aprile:2017iyp}.

This work is organized as follows: In Section~\ref{sec:CSI} we introduce the $B-L$ extension of the SM with classical conformal symmetry, where the hidden scalar could be stabilized depending on its $B-L$ charge. In this section we consider also the limits from vacuum stability and perturbativity as well as the current collider constraints on $Z'$ boson. The cosmological symmetry breaking history and the GWs generated during the phase transition are investigated in Section~\ref{sec:PTGW}. The impact of the hidden scalar on resonant leptogenesis is studied in the Section~\ref{sec:leptogenesis}. The  relic abundance of the hidden scalar in the $U(1)_{B-L}$ model is explored in Section~\ref{relic}, where we also comment briefly on the collider search of the DM particle, before we conclude in Section~\ref{conc}. The renormalization group equations (RGEs), the (reduced) cross sections for leptogenesis and DM annihilation are collected in the appendices.

\section{The conformal $U(1)_{B-L}$ model}
\label{sec:CSI}

\subsection{The basic setup}

\begin{table}[t]
  \centering
  \caption{Particle content of the conformal $U(1)_{B-L}$ model: In addition to the SM particles,  there are three RHNs $N_i$ ($i=1,2,3$), a complex singlet scalar $\Phi$ and another complex singlet scalar ${\cal S}$.}
  \label{tab:content}
  \begin{tabular}{ccccc}
  \hline\hline
            & SU(3)$_c$ & SU(2)$_L$ & U(1)$_Y$ & U(1)$_{B-L}$  \\ \hline
  $ q_L^i $    & {\bf 3}   & {\bf 2}& $+1/6$ & $+1/3$  \\
  $ u_R^i $    & {\bf 3} & {\bf 1}& $+2/3$ & $+1/3$  \\
  $ d_R^i $    & {\bf 3} & {\bf 1}& $-1/3$ & $+1/3$  \\ \hline
  $ \ell^i_L$    & {\bf 1} & {\bf 2}& $-1/2$ & $-1$  \\
  $ N_i$   & {\bf 1} & {\bf 1}& $ 0$   & $-1$  \\
  $e_R^i  $   & {\bf 1} & {\bf 1}& $-1$   & $-1$  \\ \hline
  $ H$         & {\bf 1} & {\bf 2}& $+1/2$  &  $ 0$  \\
  $ \Phi$      & {\bf 1} & {\bf 1}& $  0$  &  $+2$  \\
  ${\cal S}$      & {\bf 1} & {\bf 1}& $  0$  &  $n_x$  \\
  \hline\hline
  \end{tabular}
\end{table}

The particle content of the conformal $U(1)_{B-L}$ model is presented in Table~\ref{tab:content}, where the $q_L$, $u_R$ and $d_R$ are the SM quark doublets and singlets, $\ell_L$ and $e_R$ the SM lepton doublets and singlets, and $H$ is the SM-like Higgs doublet. Three RHNs $N_i$, a complex singlet scalar $\Phi$ with $B-L$ charge of 2 and a complex singlet scalar ${\cal S}$ with $B-L$ charge of $n_x$ are introduced to the model. To implement the EWSB, the most general scalar potential for the fields $H$ and $\Phi$ reads, which is classically scale invariant,
\begin{equation}
\label{potentialcoupled4}
V_{\rm cl}(H,\phi) \ = \
\lambda_H (H^\dagger H)^2
+ \lambda_\phi  (\Phi^\dagger \Phi)^2
-\lambda_P (H^\dagger H) (\Phi^\dagger \Phi) \,.
\end{equation}
When the scalar $\mathcal{S}$ couples to the fields $H$ and $\Phi$ via the scalar portal interactions, the full scalar potential is
\begin{equation}
\label{potentialcoupled3}
V_{\rm cl}(H,\phi,\mathcal{S})\ = \
V_{\rm cl}(H,\Phi) +
\lambda_{HS} (H^\dagger H) ({\cal S}^\dagger {\cal S}) +
\lambda_{\phi S} (\Phi^\dagger \Phi) (\mathcal{S}^\dagger {\cal S}) +
\lambda_{S} (\mathcal{S}^\dagger {\cal S})^2 \,.
\end{equation}
After spontaneous symmetry breaking, the $\Phi$ and $H$ fields develop non-vanishing VEVs, which are respectively
\begin{eqnarray}
\langle \Phi \rangle = \frac{1}{\sqrt2} v_{BL} \,, \quad
\langle H^0 \rangle = \frac{1}{\sqrt2} v_{\rm EW} \,,
\end{eqnarray}
with $H^0$ the neutral component of the doublet $H$. For simplicity, all the coupling coefficients in the potential~\eqref{potentialcoupled3} are assumed to be positive. In addition, the positivity of $\lambda_{HS}$ and $\lambda_{\phi S}$ ensures that no VEV is generated for the hidden scalar $\mathcal{S}$, which is a necessary condition for ${\cal S}$ to be a DM candidate. In the unitarity gauge, we have the following physical scalars $H=(0\;, h/\sqrt2)$, $\Phi= \phi/\sqrt2$,  $\mathcal{S}$ and its complex conjugate ${\cal S}^\dagger$. With the $B-L$ charge of $2$, the $\Phi$ scalar can give masses to the RHNs, through the $Y_\phi$ Yukawa interactions below
\begin{eqnarray}
{\cal L}_{\rm Yukawa} \ \supset \
Y_D \bar{\ell} H N + \frac12 Y_\phi \overline{N^C} \Phi N + \hbox{H.c.} \,,
\label{eq:type1}
\end{eqnarray}
where we do not show explicitly the flavor indices for the sake of clarity, and $C$ is the charge conjugate operator. In Eq.~(\ref{eq:type1}), the $Y_D$ term is responsible for the Dirac neutrino mass matrix, and the tiny neutrino masses are generated through the type-I seesaw mechanism $ m_\nu=- Y_D m_N^{-1} Y_D^{\sf T} v_{\rm EW}^2/2$, with $M_N = Y_\phi v_{BL}/\sqrt2$ the RHN mass matrix.

%~\eqref{potentialcoupled3} describe the general renormalisable gauge-invariant scalar potential
%for the three classically massless scalars as required by classical scale invariance.

When the 1-loop corrections are taken into consideration, the effective potential for the $\phi$ field is
\begin{equation}
\label{eqn:V1}
V_1(\phi;\,\mu) \ = \
\frac{\lambda_\phi(\mu)}{4}\phi^4 +\frac{\beta_{\lambda_\phi}}{8}\phi^4\left(\log\frac{\phi^2}{\mu^2}-\frac{25}{6}\right)
-\frac{\lambda_{\rm P}(\mu)}{4} h^2 \phi^2 \,,
\end{equation}
where the couplings $\lambda_\phi$ and $\lambda_P$ depend on the energy scale $\mu$, and the exact expression for the coefficient $\beta_{\lambda_\phi}$ is given in Eq.~(\ref{eqn:beta2}). Minimizing the potential in Eq.~(\ref{eqn:V1}) at the scale $\mu=v_{BL}$ gives the matching condition for the couplings; and expanding the terms in Eq.~(\ref{eqn:V1}) around the vacuum at $v_{BL}$  determines the mass of the Coleman-Weinberg field $\phi$, giving rise to
\begin{eqnarray}
\label{eqn:lambdaphi}
\lambda_\phi (v_{BL}) \ = \ \frac{11}{6} \beta_{\lambda_\phi} ,
\end{eqnarray}
and the potential can be simplified to be
\begin{equation}
V_1(\phi;\,v_{BL}) \ = \
\beta_{\lambda_\phi} \phi^4 \left[ 2\log\left(\frac{\phi^2}{v_{BL}^2}\right)-1 \right] \,.
\end{equation}
After symmetry breaking at the $U(1)_{BL}$ scale we obtain the following mass for the $U(1)_{B-L}$ gauge boson $Z'$
\begin{eqnarray}
\label{eqn:MZprime}
M_{Z'} \ = \ 2g_{BL} v_{BL} \,,
\end{eqnarray}
where $g_{BL}$ is the gauge coupling for the $U(1)_{B-L}$ gauge group. The $\phi$ field mass is therefore given by
\begin{equation}
\label{eq:phiEWSB}
m_\phi^2 \ = \
\beta_{\lambda_\phi} v^2_{BL} \approx
\frac{4 m_{S}^4-m_N^4+6 M_{Z'}^4}{16\pi^2 v_{BL}^2}\;,
\end{equation}
where we have applied the relation $\beta_{\lambda_\phi} \approx (96 g_{BL}^4+\lambda_{\phi S}^2-Y_{\phi}^4)/16\pi^2$, and $m_N$ and $m_S$ are respectively the masses for RHNs and the ${\cal S}$ scalar. One can see from Eq.~(\ref{eq:phiEWSB}) that correct spontaneous symmetry breaking of the $U(1)_{B-L}$ symmetry requires that $4 m_{S}^4+6 m_{Z'}^4>m_N^4$. Supposing $Y_\phi$ is much smaller than $g_{BL}$ and $\lambda_{\phi S}$ which implies that the RHNs are much lighter than the $v_{BL}$ scale, we have
\begin{equation}
m_\phi^2 \ \approx \ \frac{4 m_{S}^4+6 M_{Z'}^4}{16\pi^2 v_{BL}^2} \,.
\end{equation}
If both the contributions of $\lambda_{\phi S}$ and $Y_\phi$ to $\beta_{\lambda_\phi}$ are negligible, then
\begin{equation}
m_\phi^2 \ \approx \ \frac{6 M_{Z'}^4}{16\pi^2 v_{BL}^2} \,.
\end{equation}

A non-vanishing VEV of the $\phi$ field will generate the following mass parameters for the scalar potential in Eq.~(\ref{potentialcoupled4}), which is essential for the spontaneous EWSB,
\begin{eqnarray}
\label{eq:massParameters}
\mu_H^2 \ = \  - \frac12 \lambda_{\text{P}} v_{BL}^2 \,, \quad
\mu_S^2 \ = \ + \frac12 \lambda_{\phi S} v_{BL}^2 \,.
\end{eqnarray}
The VEV $v_{BL}$ also generates the mass term for the ${\cal S}$ field:
\begin{equation}
\label{ms2}
m_S^2 \ = \ \frac{1}{2} \lambda_{\phi S} v_{BL}^{2} \,,
\end{equation}
in the vacuum $S=0$, $\langle \phi \rangle = v_{BL}$, $\langle H^0 \rangle =v_{\rm EW}/\sqrt{2} = \sqrt{\lambda_{P}/\lambda_{H}} v_{BL}$. This relation is justified when the quartic coupling $\lambda_{HS}$ is sufficiently small, and therefore the EWSB contribution to $m_S^2$, i.e. the $\lambda_{HS}v^2$ term, is negligible. Here we stress that the $\mu_S^2$ term can also be negative and thus one can expect a local minimum in the direction of $S$.
The expressions for the electroweak VEV $v_{\rm EW}$ and the Higgs mass $m_h$
are analogous to the SM case.

% \begin{figure}[!htp]
%\centering
%\includegraphics[width=0.4\columnwidth]{eBL03lHs034y123105.pdf}
%\includegraphics[width=0.4\columnwidth]{eBL03lHs034y12305.pdf}
%\includegraphics[width=0.4\columnwidth]{eBL01lHs034y123105.pdf}
%\includegraphics[width=0.4\columnwidth]{eBL03lHs01y123105.pdf}
%\caption{RG evolution in CSI ESM theories with E = U(1)BBL + s(x), the initial values for those coupling are listed in the appendix B. }\label{fig:RGE}
%\end{figure}

\subsection{Limits from vacuum stability and perturbativity}
\label{sec:stability}

\begin{figure}[!htp]
  \centering
  \includegraphics[width=0.5\columnwidth]{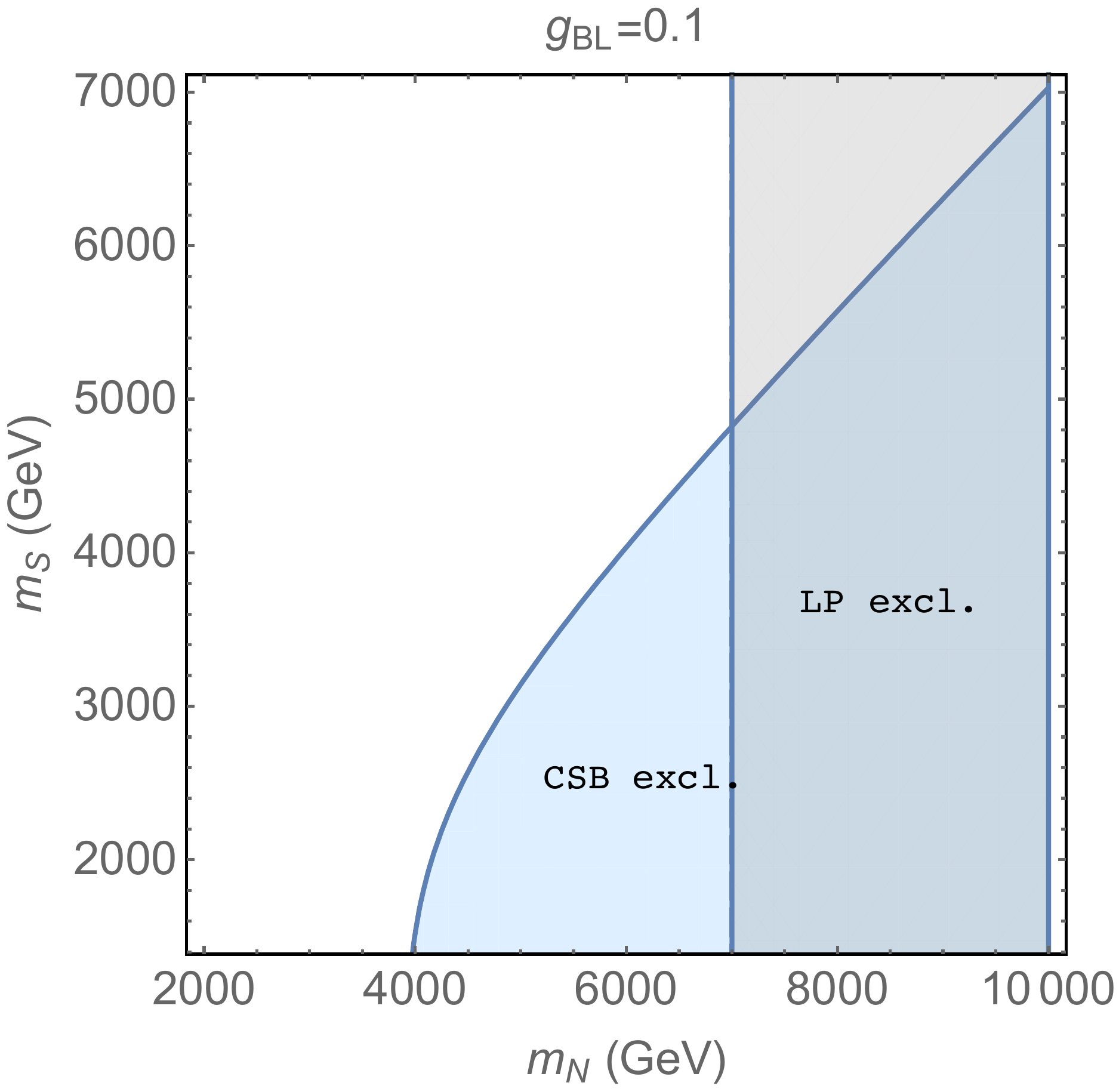}
  \caption{\label{excl} The Landau pole and vacuum stability excluded regions for the $g_{BL}=0.1$ scenario given in Fig.~\ref{fig:gw}.}
\end{figure}

For the sake of completeness we check the limits on the conformal $U(1)_{B-L}$ model from vacuum stability and perturbativity. The one-loop RGEs for all the quartic, Yukawa and gauge couplings are collected in Appendix~\ref{sec:RGE}, and the tree-level stability conditions are given as below, which is consistent with that given in Ref.~\cite{Biswas:2016ewm}:
\begin{eqnarray}
&&\lambda_H \geq 0 \,, \quad
\lambda_\phi \geq 0 \,, \quad
\lambda_{S} \geq 0 \,, \nonumber \\
&& 2\sqrt{\lambda_H\lambda_\phi} - \lambda_{P} \geq 0 \,, \quad
\lambda_{HS} - 2\sqrt{\lambda_H\,\lambda_{S}} \geq 0 \,, \quad
\lambda_{\phi S} - 2\sqrt{\lambda_\phi\,\lambda_{S}} \geq 0 \,, \nonumber \\
&&\sqrt{-\lambda_{p}+2\sqrt{\lambda_H\,\lambda_\phi}}\sqrt{\lambda_{HS}+
2\sqrt{\lambda_H\,\lambda_{S}}}
\sqrt{\lambda_{\phi S}+2\sqrt{\lambda_\phi\,\lambda_{S}}} \nonumber \\
&&+ 2\,\sqrt{\lambda_H \lambda_\phi \lambda_{S}} - \lambda_{p} \sqrt{\lambda_{S}}
+ \lambda_{HS} \sqrt{\lambda_\phi} + \lambda_{\phi S} \sqrt{\lambda_H} \geq 0 \,.
\end{eqnarray}
From these equations and relations we can find the landau pole and vacuum stability bounds on the quartic scalar couplings, the $U(1)_{B-L}$ gauge couplings $g_{BL}$,  and the $B-L$ charge $n_x$ of the hidden scalar ${\cal S}$. With the initial conditions for all SM couplings at the SM scale $\mu=m_t$ (with $m_t$ the top quark mass), we run all the couplings up to the Planck scale $\mu=M_{\rm Pl} = 1.22 \times 10^{19}$ GeV using the RGEs.

We study three different scenarios of the $U(1)_{B-L}$ which is shown in  Fig.~\ref{fig:gw} and to be compared to the current LHC constraints on the $Z'$ boson and futre GW prospects. It turns out that the vacuum stability issue is not going to be much better than in the SM, since the GW prospects of the conformal $U(1)_{B-L}$ model prefer samll quartic couplings of $\lambda_{HS,\,P}$, and a large Yukawa coupling $Y_\phi$ for the RHNs would result in Landau pole problem, since they tend to dominate the running of the quartic couplings at sufficiently high scale. The Landau pole appears at a scale much lower than $M_{\rm Pl}$ for both the second and third benchmark scenarios in Fig.~\ref{fig:gw}; as a comparison, the first scenario is much better, benefitting from a smaller coupling $g_{BL} = 0.1$. The vacuum stability and Landau pole limits on the RHN mass $m_N$ and the DM mass $m_S$ is show in Fig.~\ref{excl}, where we have set $n_x = 1$.

\subsection{Current $Z'$ limits}
\label{sec:collider}

\begin{figure}[]
  \centering
  \includegraphics[width=0.52\textwidth]{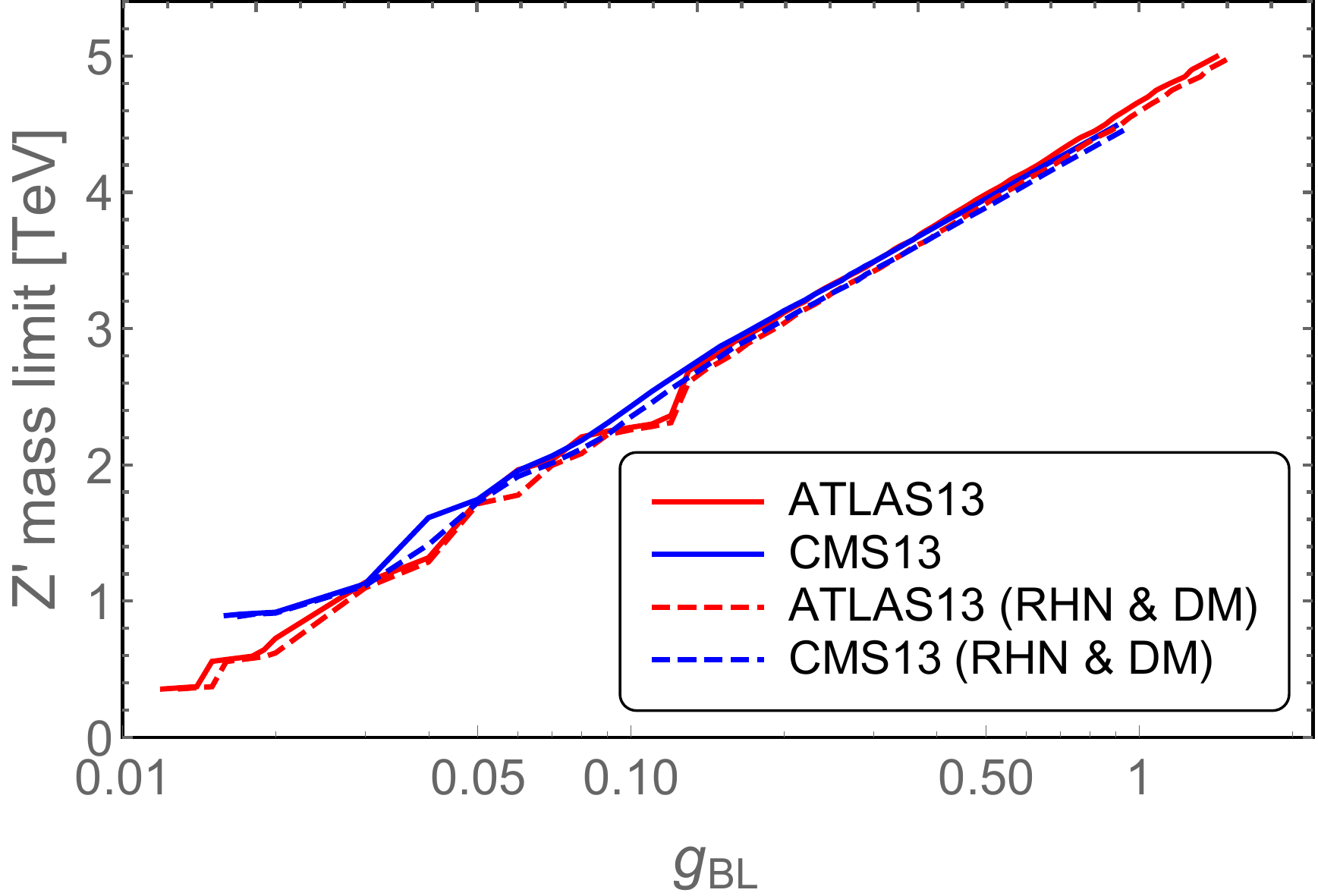}
  \caption{Dilepton limits on the $Z'$ boson mass from the 13 TeV data by ATLAS~\cite{ATLAS:2016cyf} (red) and CMS~\cite{CMS:2016abv} (blue), as function of the gauge coupling $g_{BL}$. The solid curves assume that the $Z'$ boson decays only into the SM fermions, while for the dashed curves $Z'$ decays also into the three RHNs and DM.}
  \label{fig:Zprime}
\end{figure}

For a TeV-scale $v_{BL}$, the $Z'$ mass is stringently constrained by the dilepton data $pp \to Z' \to \ell^+ \ell^-$ (with $\ell = e,\, \mu$) at the LHC~\cite{Patra:2015bga, Lindner:2016lpp}. For a sequential $Z'$ boson with the same couplings as in the SM, the current ATLAS and CMS 13 TeV data requires that $M_{Z'} > 4.05$ TeV at the 95\% confidence level~\cite{ATLAS:2016cyf, CMS:2016abv}. The production cross section $\sigma (pp \to Z' \to \ell^+ \ell^-)$ in the $U(1)_{B-L}$ model can be obtained by rescaling that of a sequential heavy $Z'$ boson, as function of the gauge coupling $g_{BL}$~\cite{Dev:2016xcp}. To this end, the partial decay widths of the $Z'$ boson into the SM fermions, the heavy RHNs and the scalar ${\cal S}$ are respectively
\begin{eqnarray}
\Gamma (Z' \to f\bar{f}) & \ = \ &
\frac{S_f N_C^f (B_f-L_f)^2 g_{BL}^2 M_{Z'}}{48\pi} \,, \nonumber \\
\Gamma (Z' \to NN) & \ = \ &
\frac{g_{BL}^2 M_{Z'}}{96\pi}
\left( 1 - \frac{4m_N^2}{M_{Z'}^2} \right)^{3/2} \,, \nonumber \\
\Gamma (Z' \to {\cal S} {\cal S}^\dagger) & \ = \ &
\frac{n_{x}^2 g_{BL}^2 M_{Z'}}{192\pi}
\left( 1 - \frac{4m_{\cal S}^2}{M_{Z'}^2} \right)^{3/2} \,,
\end{eqnarray}
with $N_C$ the color factor (3 for quarks and 1 otherwise), $B_f$ and $L_f$ the baryon and lepton numbers for the SM fermions, $S_f = 1$ for the quarks and charged leptons and $1/2$ for the light neutrinos. All these decay modes are universally proportional to the gauge coupling $g_{BL}$. In the absence of the heavy RHNs and the ${\cal S}$ field, the branching fraction ${\rm BR} (Z' \to \ell^+ \ell^-)$ is a constant, being $8/23$, in the limit of $M_{Z'} \gg m_f$, and the production cross section $\sigma (pp \to Z') \propto g_{BL}^2$. As a result, when $g_{BL}$ gets larger, the dilepton limits on the $Z'$ mass tend to be stronger. The constraints from the ATLAS~\cite{ATLAS:2016cyf} and CMS~\cite{CMS:2016abv} 13 TeV data are shown respectively as the solid red and blue curves in Fig.~\ref{fig:Zprime}. As a comparison, we also show in Fig.~\ref{fig:Zprime} the dilepton limits in the presence of the three RHNs and DM as the dashed curves, assuming their masses are significantly lower than $M_{Z'}/2$ thus the decays $Z' \to NN,\, {\cal S} {\cal S}^\dagger$ are kinematically allowed. As a result of these extra decays modes, the dilepton limits in Fig.~\ref{fig:Zprime} become slightly weaker. For illustration purpose, we adopt three different benchmark values of $g_{BL} = 0.1$, $0.3$ and $1.0$, interpret the solid lines in Fig.~\ref{fig:Zprime} and obtain the current dilepton constraints on the $Z'$ boson mass and the corresponding limits on $v_{BL}$, which are collected in Table~\ref{tab:limits}.
%For a $g_{BL} = 1$ in the $U(1)_{B-L}$ model, the dilepton limits go up to 4.66 (4.59) TeV in the absence (presence) of the decay modes into RHNs and DM; for a smaller $g_{BL} = 0.1$, the dilepton constraints are much weaker, being only 2.42 (2.35) TeV.
At the high-luminosity LHC (HL-LHC) and future 100 TeV colliders, the prospects of the $Z'$ boson could be largely improved~\cite{Diener:2010sy, Godfrey:2013eta, Rizzo:2014xma}.

\begin{table}[!t]
  \centering
  \caption{\label{tab:limits} The lower bounds on the $Z'$ boson mass $M_{Z'}$ and the $v_{BL}$ scale in the $U(1)_{B-L}$ model from the current LHC13 data~\cite{ATLAS:2016cyf, CMS:2016abv} (cf. Fig.~\ref{fig:Zprime}). }
  \begin{tabular}{l c c c c c}
  \hline\hline
  & \multicolumn{2}{c}{without RHNs \& DM} && \multicolumn{2}{c}{with RHNs \& DM} \\ \cline{2-3} \cline{5-6}
  $g_{BL}$ & $M_{Z'}$ [TeV] & $v_{BL}$ [TeV] && $M_{Z'}$ [TeV] & $v_{BL}$ [TeV] \\ \hline
  0.1 & $2.42$ & $17.2$ && $2.35$ & $16.6$  \\
  0.3 & $3.49$ & $8.22$ && $3.43$ & $8.08$ \\
  1.0 & $4.66$ & $3.30$ && $4.59$ & $3.25$ \\ \hline
  \end{tabular}
\end{table}

\section{Phase transition dynamics and gravitational wave signatures}
\label{sec:PTGW}

As the Universe cools down, the EWSB is induced by the dynamical breaking of the $U(1)_{B-L}$, i.e., phase transition. If the phase transition is strong first order, GWs can be produced and potentially probed by the space-based interferometers like LISA.

\subsection{Dynamical $U(1)_{B-L}$ breaking and phase transition}

In this section, we first demonstrate the calculation of the phase transition, which is determined by the thermal potential.
The finite temperature corrections to the effective potential at one loop are given by
\begin{equation}
V_{1}(\phi, T)=\frac{T^{4}}{2 \pi^{2}} \sum_{i} n_{i} J_{B, F}\left(\frac{M_{i}^{2}(\phi)}{T^{2}}\right)
\end{equation}
where the functions $J_{B, F}(y)$ are
\begin{equation}
J_{B, F}(y)=\pm \int_{0}^{\infty} d x x^{2} \ln \left[1 \mp \exp \left(-\sqrt{x^{2}+y}\right)\right]
\end{equation}
with the upper (lower) sign corresponding to bosonic (fermionic) contributions.
Here, in order to describe the high-$T$ and low-$T$ behaviors appropriately, the above integrals $J_{B, F}$ can be expressed as a sum of the second kind of modified Bessel functions $K_2(x)$~\cite{Bernon:2017jgv},
\begin{equation}
J_{B, F}(y)=\lim _{N \rightarrow+\infty} \mp \sum_{l=1}^{N} \frac{( \pm 1)^{l} y}{l^{2}} K_{2}(\sqrt{y} l)\;.
\end{equation}

%With the hidden $B-L$ charged scalar effects being taken into account, the first approach being adopted by Ref.~\cite{Hambye:2018qjv} are
%\begin{equation}
%V^A_{1}(\phi, T)= \frac{2T^4}{2\pi^2}J_B(\frac{m_S^2}{T^2})+\Pi_\phi \phi^2+3\frac{T^4}{2\pi^2}J_B(\frac{m^2_{Z'}}{T^2})\;.
%\end{equation}
%Drop the second terms of $V^A_{1}(\phi, T)$ as

%adopted by Ref.~\cite{1604.05035} for the study of classical scale invariant $B-L$ phase transition, one have,
% \begin{equation}
%V^B_{1}(\phi, T)= \frac{2T^4}{2\pi^2}J_B(\frac{m_S^2}{T^2})+3\frac{T^4}{2\pi^2}J_B(\frac{m^2_{Z'}}{T^2})\;.
%\end{equation}
The dominant contributions come from the hidden scalar ${\cal S}$, RHNs $N_i$ and the extra gauge field $Z'$.
% the thermal potential is given by
% \begin{equation}
%V_{1}(\phi, T)= \frac{2T^4}{2\pi^2}J_B(\frac{m_S^2+\Pi_S}{T^2})+\frac{T^4}{2\pi^2}J_B(\frac{m_N^2}{T^2})+3\frac{T^4}{2\pi^2}J_B(\frac{m^2_{Z'}+\Pi_{Z'}}{T^2})\;.
%\end{equation}
%%When the right handed neutrino contribution is negligible, one have
%% \begin{equation}
%%V^B_{1}(\phi, T)= \frac{2T^4}{2\pi^2}J_B(\frac{m_S^2+\Pi_S}{T^2})+3\frac{T^4}{2\pi^2}J_B(\frac{m^2_{Z'}+\Pi_{Z'}}{T^2})\;.
%%\end{equation}
%%Suppose the hidden scalar contributions are negligible rather than the right handed neutrinos, one have,
%% \begin{equation}
%%V^C_{1}(\phi, T)=\frac{T^4}{2\pi^2}J_B(\frac{m_N^2}{T^2})+3\frac{T^4}{2\pi^2}J_B(\frac{m^2_{Z'}+\Pi_{Z'}}{T^2})\;.
%%\end{equation}
%When both the hidden scalar contributions and the right handed neutrino contributions are all negligible, which reduce to the scale invariant $B-L$ case, with
% \begin{equation}
%V^{BL}_{1}(\phi, T)=3\frac{T^4}{2\pi^2}J_B(\frac{m^2_{Z'}+\Pi_{Z'}}{T^2})\;.
%\end{equation}
The field dependent mass and thermal corrections are given respectively by
\begin{eqnarray}
&&m_S^2 \ = \ \frac{\lambda_{\phi S}}{2}\phi^2 \,, \quad
m^2_{Z'} \ = \ 4 g_{B-L}^2 \phi^2 \,,\\
&&\Pi_{Z'} \ = \ 4g_{B-L}^2T^2 \,, \quad
\Pi_S \ = \ (g_{B-L}^2+\frac{\lambda_{\phi S}}{12})T^2 \,, \quad
\Pi _\phi=(\frac{\lambda_{\phi s}}{12}+g_{B-L}^2+Y_\phi^2)T^2 \,.
\end{eqnarray}

%We present in Fig.~\ref{fig:CSIVCTC} the potential shape with the Universe cools down, where the $T_C$ is the temperature where one have the potential degeneracy behavior. For $T>T_C$, one have the $U(1)_{B-L}$ symmetry being restored. And when the temperature
%drops below $T_C$, one may have the $U(1)_{B-L}$ symmmetry broken after the bubble nucleation occurs at $T_n$. The four figures demonstrate the roles played by the hidden scalars and right handed heavy neutrinos, that can pulls down the critical temperature $T_C$.
%The $BP-D$ reflect the situation of scale invariant $B-L$ model.
%
%\begin{figure}[t]
%  \centering
%    \includegraphics[width=0.4\columnwidth]{CSI_EWPTBPA.pdf}
%      \includegraphics[width=0.4\columnwidth]{CSI_EWPTBPB.pdf}
%  \includegraphics[width=0.4\columnwidth]{CSI_EWPTBPC.pdf}
%    \includegraphics[width=0.4\columnwidth]{CSI_EWPTBPD.pdf}
%\caption{\label{fig:CSIVCTC}
%The phase evolution process with the Universe cools down described by the thermal potential $V^{A,B,C,D}$ with the parameters been chosen as: $m_{Z^\prime}=4.59$ TeV, $m_S=3.0$ TeV, $v_{BL}=3.25$ TeV, and $m_N=4.6$ TeV.}
%\end{figure}

\subsection{Gravitational waves signals}

The bounce configuration of the nucleated bubble, i.e. the bounce configuration of the field that connects the $U(1)_{B-L}$ broken vacuum (true vacuum) and the false vacuum (here it can be $U(1)_{B-L}$ conserving vacuum), can be obtained by extremizing
\begin{eqnarray}
S_3(T) \ = \ \int 4\pi r^2d r\bigg[\frac{1}{2}\big(\frac{d \phi_b}{dr}\big)^2+V(\phi_b,T)\bigg]
\end{eqnarray}
 through solving the equation of motion for $\phi_b$ (it is $\phi$ for the scenario under study),
\begin{eqnarray}
\frac{d^2\phi_b}{dr^2}+\frac{2}{r}\frac{d\phi_b}{dr}-\frac{\partial V(\phi_b)}{\partial \phi_b} \ = \ 0\;,
\end{eqnarray}
with the boundary conditions of
\begin{eqnarray}
\lim_{r\rightarrow \infty}\phi_b \ = \ 0 \,, \quad
\frac{d\phi_b}{d r}|_{r=0} \ = \ 0\;.
\end{eqnarray}
At the nucleation temperature $T_n$, the thermal tunneling probability for bubble nucleation per horizon volume and per horizon time is of order unity with~\cite{Affleck:1980ac,Linde:1981zj,Linde:1980tt},
\begin{eqnarray}
\Gamma \ \approx \ A(T)e^{-S_3/T} \ \sim \ 1\;.
\end{eqnarray}

Two parameters are crucial for the calculations of GW emission:
\begin{itemize}
  \item The parameter $\alpha$. It describes the strength of the phase transition, and is defined as the energy density released from the strong first order EWPT normalized by the total radiation energy density $\rho_R = \pi^2 g_\star T_\star^4/30$:
      \begin{eqnarray}
      \alpha=\frac{\Delta\rho}{\rho_R}\;,
      \end{eqnarray}
      where $\Delta\rho$ is the latent heat released in phase transition, i.e. the difference of the energy density between the false and the true vacuum.
  \item  The parameter $\beta$. It describes roughly the inverse time duration of the strong first order EWPT, and characterizes the GW spectrum peak frequency, which is connected with the action $S_3$ through
\begin{eqnarray}
\frac{\beta}{H_n}=T\frac{d (S_3(T)/T)}{d T}|_{T=T_n}\; ,
\end{eqnarray}
where $H_n$ is the Hubble parameter at the bubble nucleation temperature $T_n$.
\end{itemize}

We are now ready to calculate the stochastic GW background generated during the first order phase transition. Significant progress has been made in recent years on the calculations of the GW from phase transitions (see e.g. Ref.~\cite{Caprini:2015zlo,Cai:2017cbj,Weir:2017wfa} for recent reviews). It is now generally believed that the dominant source for the GW production in this process is the sound waves (SWs) in the plasma which lasts long after the phase transition completes~\cite{Hindmarsh:2013xza,Hindmarsh:2015qta}, though the bubble collision contribution has also been theoretically well modeled~\cite{Jinno:2017fby,Jinno:2017ixd, Jinno:2016vai,Cutting:2018tjt,Kosowsky:1991ua, Kosowsky:1992rz,Kosowsky:1992vn,Huber:2008hg}. Another contribution comes from the Magnetohydrodynamic (MHD) turbulence in the magnetized plasma with high Reynolds number~\cite{Caprini:2009yp,Binetruy:2012ze}. The total resultant energy density spectrum can be approximated by the following linear summation of the individual contributions above:
\begin{equation}
\Omega_{\text{GW}}h^{2} \ \simeq \ \Omega_{\rm col}h^{2}+\Omega_{\rm sw}h^{2}+\Omega_{\rm turb}h^{2} ,
\end{equation}
and we neglect in the following the contribution from bubble collision $\Omega_{\rm col}$.

The GW spectrum from the dominant SWs can be found by fitting to the result of numerical simulations with
the fluid-scalar field model~\cite{Hindmarsh:2015qta}:
\begin{eqnarray}
\Omega_{\textrm{sw}}h^{2} & \ = \ &
2.65\times10^{-6}\left( \frac{H_{n}}{\beta}\right)^{2}\left(\frac{\kappa_{v} \alpha}{1+\alpha} \right)^{2}
\left( \frac{100}{g_{\ast}}\right)^{1/3} \nonumber \\
&&\times v_{w} \left(\frac{f}{f_{sw}} \right)^{3} \left( \frac{7}{4+3(f/f_{\textrm{sw}})^{2}} \right) ^{7/2} \ ,
\label{eq:soundwaves}
\end{eqnarray}
where $g_{\ast}$ is the relativistic degrees of freedom in the plasma at the time of EWPT and $f_{\text{sw}}$ is the present peak frequency of the spectrum:
\begin{equation}
f_{\textrm{sw}} \ = \ 
1.9\times10^{-5}\frac{1}{v_{w}}\left(\frac{\beta}{H_{n}} \right) \left( \frac{T_{n}}{100\textrm{GeV}} \right) \left( \frac{g_{\ast}}{100}\right)^{1/6} \textrm{Hz} \,.
\end{equation}
In addition, the factor $\kappa_{v}$ is the fraction of latent heat transformed into the kinetic energy of the fluid and
can be found by solving the hydrodynamic velocity profiles of the bubbles~\cite{Espinosa:2010hh,Alves:2018jsw,Alves:2018oct}.
\begin{figure}[t]
  \centering
  \includegraphics[width=0.55\columnwidth]{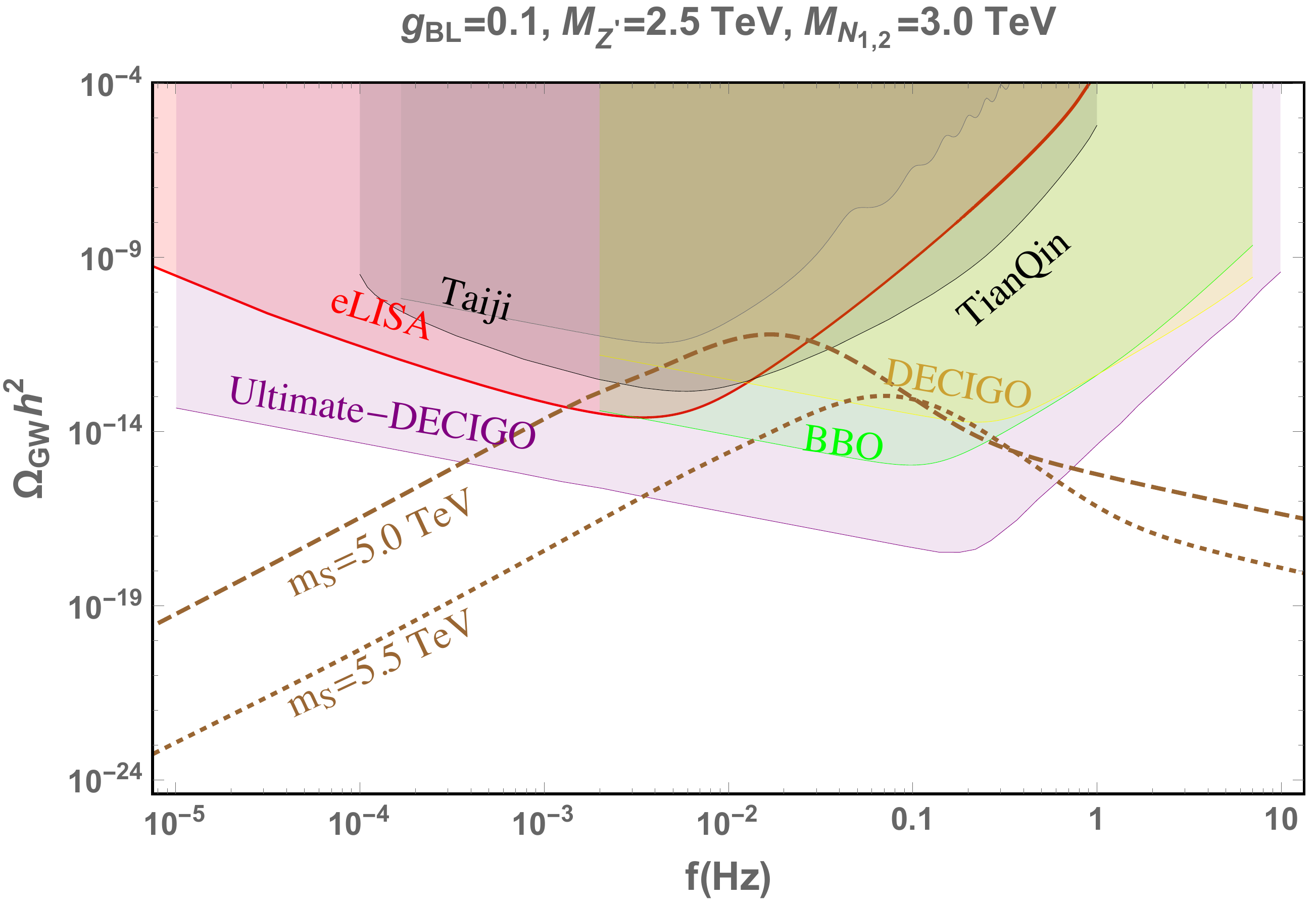}
  \includegraphics[width=0.55\columnwidth]{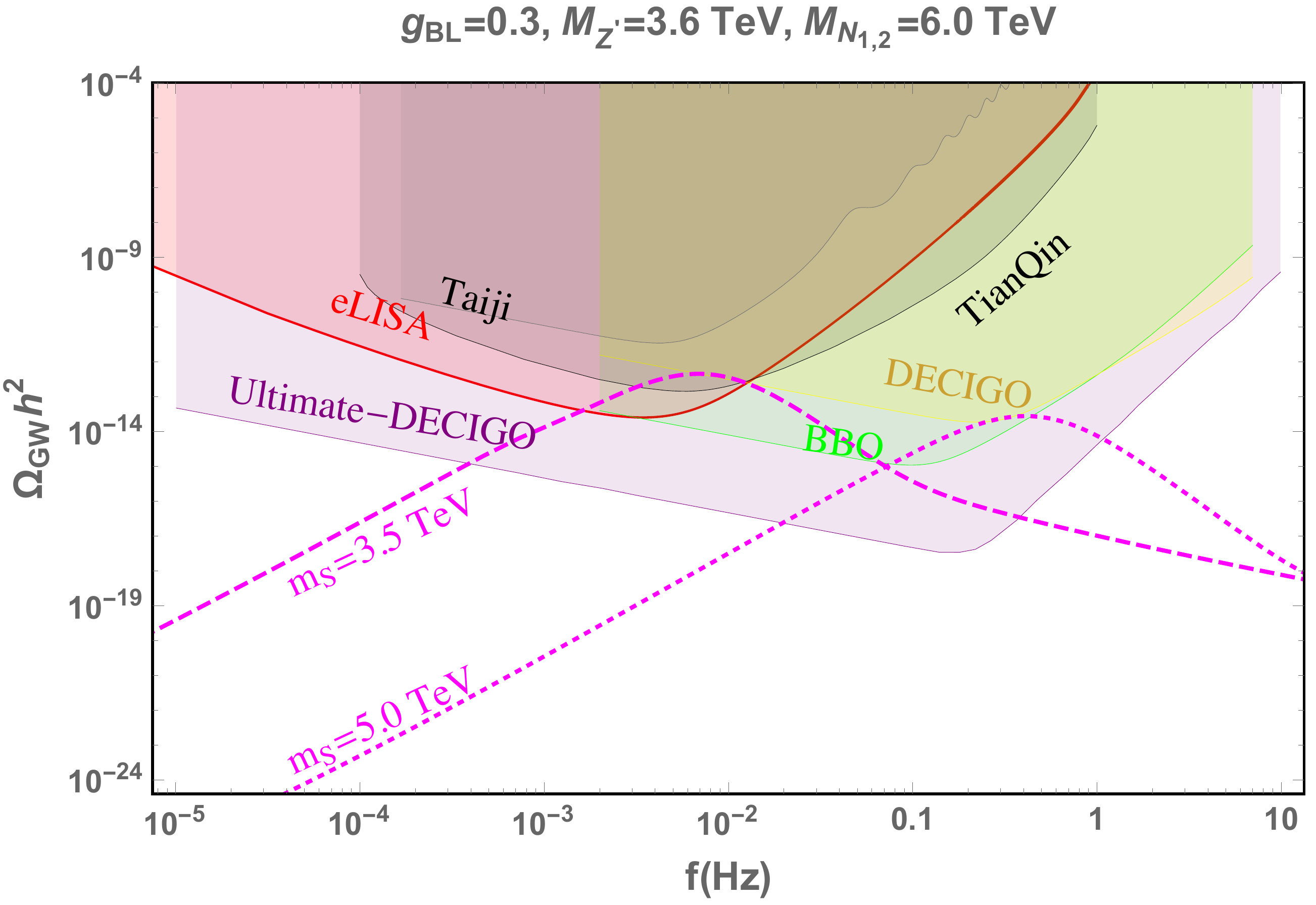}
  \includegraphics[width=0.55\columnwidth]{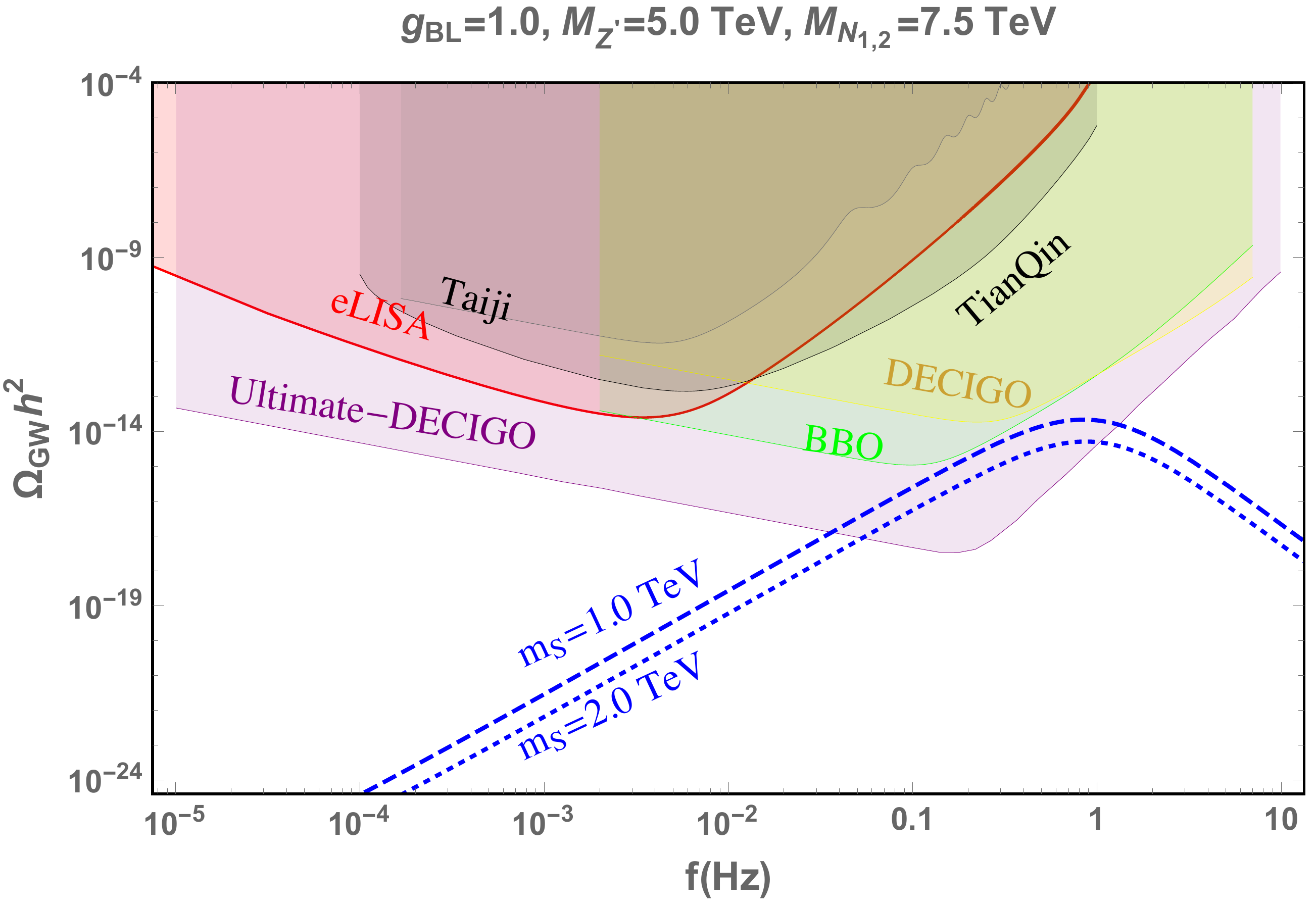}
  \caption{\label{fig:gw} The expected GW spectra for three benchmark scenarios in the conformal $U(1)_{B-L}$ model, shown as the dashed curves, with respectively $g_{BL} = 0.1$ (top), $0.3$ (middle) and $1$ (bottom). The shaded regions demonstrate the prospects for different GW experiments.}
\end{figure}

The GW spectrum from the MHD turbulence can be theoretically modelled with inputs of the magnetic and turbulence power spectra~\cite{Kosowsky:2001xp,Caprini:2009yp,Gogoberidze:2007an,Niksa:2018ofa} and improved by numerically evolving the MHD equations~\cite{Pol:2019yex,Brandenburg:2017neh}. A fitting formula is also available~\cite{Caprini:2009yp,Binetruy:2012ze}:
\begin{eqnarray}
\Omega_{\textrm{turb}}h^{2} & \ = \ & 
3.35\times10^{-4}\left( \frac{H_{n}}{\beta}\right)^{2}\left(\frac{\kappa_{turb} \alpha}{1+\alpha} \right)^{3/2} \left( \frac{100}{g_{\ast}}\right)^{1/3}  \nonumber \\
  && \times v_{w}\frac{(f/f_{\textrm{turb}})^{3}}
  {[1+(f/f_{\textrm{turb}})]^{11/3}(1+8\pi f/h_{\ast})} .
\label{eq:mhd}
\end{eqnarray}
Here the peak frequency $f_{\rm turb}$ is given by
\begin{equation}
f_{\textrm{turb}} \ = \ 2.7\times10^{-5}\frac{1}{v_{w}}\left(\frac{\beta}{H_{n}} \right) \left( \frac{T_{n}}{100\textrm{GeV}} \right) \left( \frac{g_{\ast}}{100}\right)^{1/6} \textrm{Hz} \,.
\end{equation}
The energy fraction tranferred to the MHD turbulence $\kappa_{\text{turb}}$ is uncertain as of now and can vary
between $5\%$ to $10\%$ of $\kappa_v$~\cite{Hindmarsh:2015qta}. Here we take tentatively $\kappa_{\text{turb}} = 0.1 \kappa_v$.

Summing up the results in Eqs.~(\ref{eq:soundwaves}) and (\ref{eq:mhd}), we can obtain the total GW energy density spectrum. The expected GW energy spectra for three benchmark scenarios with different $g_{BL}$ values are shown in Fig.~\ref{fig:gw}. The color-shaded regions on the top are the experimentally sensitive regions for several proposed space-based GW detectors. As discussed earlier, the GW signal comes mainly from SWs. Our study indicates that increasing the RHN masses may leads to a decrease of the phase transition temperature, while its mass is severely bounded by the EWSB conditions given in Eq.~(\ref{eq:phiEWSB}).  The three panels of Fig.~\ref{fig:gw} demonstrate that the amplitudes of GW signal spectra decrease as $g_{BL}$ increases, which implies that the GW prospects are weaker when the $B-L$ charge of DM scalar is large. The hidden scalar ${\cal S}$ is useful for generating the proper vacuum barrier at
the nucleation temperature. Furthermore, a larger hidden scalar mass leads to a lower GW amplitude and a higher peak frequency for the GW spectrum.

%\begin{figure}[t]
%  \centering
%  \includegraphics[width=0.7\columnwidth]{BM-SNR.pdf}
%\caption{\label{fig:snr}
%The SNR of the gravitational wave signals versus $m_{h_2}$ for the benchmarks shown in Table 1 for
%proposed space-based gravitational wave detectors.}
%\end{figure}

To assess the discovery prospects of the GW spectra, we calculate the signal-to-noise ratio (SNR) with the definition adopted by Ref.~\cite{Caprini:2015zlo}:
\begin{eqnarray}
\text{SNR} \ = \ \sqrt{\mathcal{T} \int_{f_{\text{min}}}^{f_{\text{max}}} df
\left[ \frac{h^2 \Omega_{\text{GW}}(f)}{h^2 \Omega_{\text{exp}}(f)}
  \right]^2} ,
\end{eqnarray}
where $h^2 \Omega_{\text{exp}}(f)$ is the experimental sensitivity for the detectors and $\mathcal{T}$ is the mission duration in unit of year for each experiment. Here we assume $\mathcal{T}=5$. For the LISA configurations with
four links, the suggested threshold SNR for discovery is 50~\cite{Caprini:2015zlo}.
For the six link configurations as drawn here, the
uncorrelated noise reduction technique can be used and the suggested SNR threshold can be as low as 10~\cite{Caprini:2015zlo}. The GW spectrum of the $m_S=5.0$ TeV case for the $g_{BL}=0.1$ and $g_{BL}=0.3$ scenarios are able to be detected by LISA, with respectively ${\rm SNR}=22$ and ${\rm SNR}=5$.

\section{Resonant Leptogenesis}
\label{sec:leptogenesis}

\begin{figure}[]
  \centering
  \begin{tabular}{ccc}
  \includegraphics[width=0.25\textwidth]{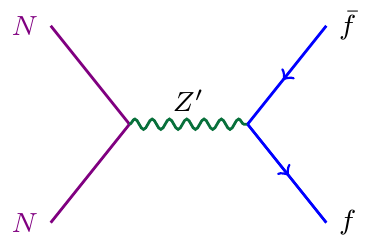} &
  \includegraphics[width=0.25\textwidth]{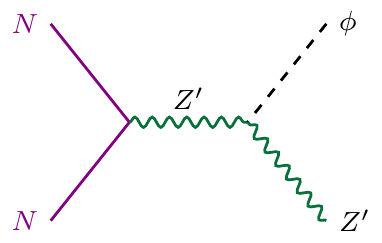} &
  \includegraphics[width=0.25\textwidth]{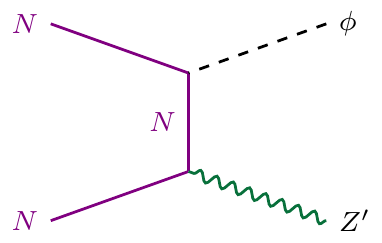} \\
  (a) & (b) & (c) \\
  \includegraphics[width=0.25\textwidth]{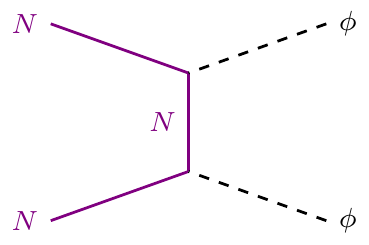} &
  \includegraphics[width=0.25\textwidth]{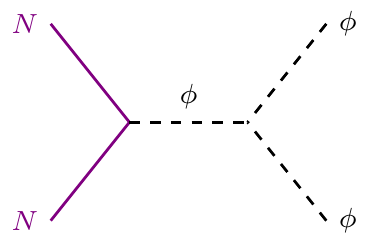} &
  \includegraphics[width=0.25\textwidth]{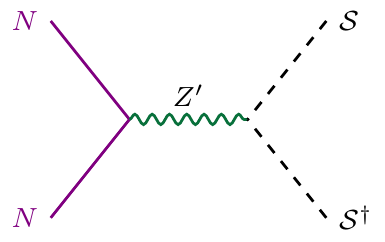} \\
  (d) & (e) & (f) \\ &
  \includegraphics[width=0.25\textwidth]{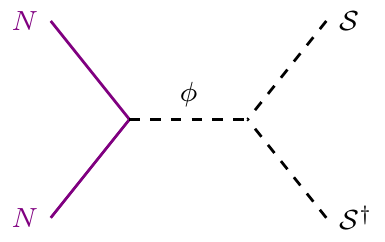} \\
  & (g)
  \end{tabular}
  \caption{Feynman diagram for the process (a) $N N \to f\bar{f}$, (b) and (c) $NN \to Z^\prime \Phi$, (d) and (e) $NN \to \Phi\Phi$, and (f) and (g) $NN \to {\cal S} {\cal S}^\dagger$.  }
  \label{fig:diagram1}
\end{figure}

For the case of leptogenesis occurring via flavor oscillations of the heavy right-handed neutrinos in the classically conformal models, we refer the readers to Ref.~\cite{Khoze:2013oga} to look for some  details. For TeV scale RHNs, it is necessary to use the resonant leptogenesis mechanism~\cite{Covi:1996wh, Flanz:1996fb, Pilaftsis:1997jf, Pilaftsis:2003gt} in order to avoid the Davidson-Ibarra bound~\cite{Davidson:2002qv}. For simplicity, we assume the two RHN mass eigenstates $N_1$ and $N_2$ are almost degenerate with the mass $m_{N}$ and a small splitting $\Delta m_N / m_N \ll 1$, and the third RHN $N_3$ significantly heavier. The heavy $Z'$ boson, the conformal scalar $\phi$ and the DM scalar ${\cal S}$ play important roles in the generation of lepton asymmetry from the decay of RHNs, as they would induce processes that dilute the heavy RHNs by two units, thus reducing the lepton and baryon asymmetry significantly in a large regions of parameter space~\cite{Blanchet:2009bu, Blanchet:2010kw, Iso:2010mv, Okada:2012fs, Heeck:2016oda, Dev:2017xry}. Such $\Delta N = 2$ processes include
\begin{eqnarray}
NN \ \to \ f\bar{f},\;\;  Z^\prime \phi,\;\; \phi\phi,\;\; {\cal S} {\cal S}^\dagger \,,
\end{eqnarray}
with $f$ running over all the flavors of SM quarks, charged leptons and neutrinos. The corresponding Feynman diagrams are shown in Fig.~\ref{fig:diagram1}. One should note that the scalar mixing between $h$ and $\phi$ however does not play any role in freeze-out leptogenesis since the latter takes place prior to EWSB. Thus the processes $NN \to f \bar{f}$ are only mediated by the $Z'$ gauge boson. The Feynman diagrams $NN \to Z' \phi$ in (b) and (c) are mediated by the gauge and Yukawa couplings, the process $NN \to \phi\phi$ in (d) and (e) by the Yukawa couplings and the scalar quartic coupling $\lambda_\phi$ in the potential (1.4). If the DM ${\cal S}$ is lighter than the RHNs $N_{1,\,2}$, i.e. $m_{\cal S} \lesssim m_N$, the process $NN \to {\cal S} {\cal S}^\dagger$ in (f) and (g) is also important in some region of the parameter space, which is induced through both the gauge ($Z'$) and scalar ($\phi$) portals. There exists in principle also the process $NN \to Z' Z'$. However, in the conformal theories the RHN masses are constrained as shown in Eq.~(\ref{eq:phiEWSB}); with $N_3$ heavier than $N_{1,2}$ and neglecting the DM mass, Eq.~(\ref{eq:phiEWSB}) implies that $m_N = m_{N_{1,2}} < 2^{1/8} M_{Z'} \simeq 1.09 M_{Z'}$. Then the process $NN \to Z'Z'$ process is insignificant, suppressed by the kinematical space.

%\begin{figure}[t!]
%  \centering
%  \begin{array}
%  \includegraphics[width=0.25\textwidth]{diagram1.pdf} &
%  \includegraphics[width=0.25\textwidth]{diagram2.pdf}
%  \end{array}
%
%\end{figure}

%In the early Universe the annihilation processes $NN \to f\bar{f}$ would

The Boltzmann equations, which govern the evolution of the RHN number density and the lepton asymmetry, are given by
\begin{eqnarray}
\label{eqn:Boltzmann}
\frac{n_\gamma H_N}{z} \frac{{\rm d} \eta_N}{{\rm d}z} \ &=& \
- \left[ \left( \frac{\eta_N}{\eta_N^{\rm eq}} \right)^2 - 1 \right] \, 2 \gamma_{NN}
- \left( \frac{\eta_N}{\eta_N^{\rm eq}} - 1 \right) \left[ \gamma_D + \gamma_{s} + 2\gamma_{t} \right] \,, \\
\label{eqn:Boltzmann2}
\frac{n_\gamma H_N}{z} \frac{{\rm d} \eta_{\Delta L}}{{\rm d}z} \ &=& \
\gamma_D \left[ \varepsilon_{\rm CP} \left( \frac{\eta_N}{\eta_N^{\rm eq}} - 1 \right) - \frac23 \eta_{\Delta L} \right]
- \frac23 \eta_{\Delta L} \left[ \frac{\eta_N}{\eta_N^{\rm eq}} \gamma_{s} + 2\gamma_{t} \right] \,,
\end{eqnarray}
where $z\equiv m_N/T$ is a dimensionless parameter, $H_N\equiv H(z=1)\simeq 17m_N^2/M_{\rm Pl}$ is the Hubble expansion rate at temperature $T=m_N$, $n_\gamma=2T^3\zeta(3)/\pi^2$  is the number density of photons, and $\eta_N\equiv n_N/n_\gamma$ is the normalized number density of RHN (similarly $\eta_{\Delta L}=(n_L-n_{\bar{L}})/n_\gamma$ for the lepton asymmetry).  The $\gamma$'s are the various thermalized interaction rates: $\gamma_D$ for the RHN decay $N \to LH$,  and $\gamma_{s} = \gamma_{H s} + \gamma_{Vs}$ and $\gamma_{t} = \gamma_{H t} + \gamma_{V t}$ the standard $\Delta L = 1$ scattering processes as in Refs.~\cite{Giudice:2003jh, Pilaftsis:2003gt}  with the subscripts $s,t$ denoting respectively the $s$ and $t$-channel exchange of the SM Higgs doublet $H$ or the SM gauge bosons $V = W_{i},\,B$ (with $i=1,2,3$) before EWSB. %for example, the process ${\phi t}$ corresponds to $N+Q_3 \to L+t_R$ through the $t$-channel exchange of the SM Higgs field $\phi$ (here $Q_3$ and $t_R$ are respectively the third-generation quark doublet and right-handed top quark in the SM).
%and $\gamma_{\phi s}$ corresponds to the process $N+\psi\to Q_3 \bar{Q}_3$ via s-channel $\phi$ exchange process and $\gamma_{A,s}$ denotes the emission of a gauge boson ($W,B$) in then final state via a s-channel exchange of $\psi$ e.g. the process $N+\phi\to \psi + W$;
Here the integration over different momenta has already been
performed, assuming implicitly kinetic equilibrium. The new scattering processes  in our model in Fig.~\ref{fig:diagram1} correspond to the scattering rates $\gamma_{NN}$, and all the corresponding reduced cross sections $\hat{\sigma} (NN \to f\bar{f},\, Z'\phi,\, \phi\phi,\, {\cal S} {\cal S}^\dagger)$  are collected in Appendix~\ref{sec:leptogenesis_xs}. The prefactor of 2 in Eq.~(\ref{eqn:Boltzmann}) accounts for the reduction of RHN by unit of two~\cite{Heeck:2016oda}.
%$NN \to HH/AA$ via t-channel $N$ exchange.
The  thermal corrections to the SM particles are included in the calculation~\cite{Giudice:2003jh, Dev:2014laa}. If the $\gamma_{NN}$ term is comparable or larger than other terms on the r.h.s. of Eq.~(\ref{eqn:Boltzmann}), these extra processes in Fig.~\ref{fig:diagram1} could significantly dilute the RHN number density before the  sphaleron decoupling temperature $T_c\simeq 131.7$ GeV~\cite{DOnofrio:2014rug}, thus potentially  making type-I seesaw freeze-out leptogenesis ineffective. Then we can set limits on the heavy particle masses and the couplings in the conformal $U(1)_{B-L}$ model.

To be concrete, we consider two distinct scenarios, i.e. without and with the DM particle ${\cal S}$ involved in the lepton asymmetry generation in the RHN decay, with the first one corresponding to the limit of $m_{\cal S} \gg m_N$ and the second one $m_{\cal S} \lesssim m_N$. In both of the two cases, the dilution effect depends on the effective neutrino mass $\widetilde{m}\equiv v^2 (Y_D^\dag Y_D)_{11}/m_N$ (or effectively on the Yukawa coupling $Y_D$) and the CP asymmetry $\varepsilon_{\rm CP}$. Since in this paper we are mostly concerned with the role of the new particles in the lepton asymmetry generation in RHN decay $N \to LH$, we will not consider the flavor structure details in the neutrino sector but fix $\widetilde{m}\simeq \sqrt{\Delta m^2_{\rm atm}} \simeq 50 \, {\rm meV}$, without any significant tuning or cancellation in the type-I seesaw formula for light neutrino masses~\cite{Dev:2017xry}. A large CP-asymmetry $\varepsilon_{\rm CP}$ can then be generated by the resonant enhancement mechanism, and go up to order one if $\Delta m_N \sim \Gamma_N$, with $\Gamma_N$ is the averaged RHN decay width~\cite{Blanchet:2009bu, Dev:2017wwc}. For the sake of concreteness, we adopt the value of $\varepsilon_{\rm CP} = 10^{-2}$ throughout this paper.

\begin{figure}[]
  \centering
  \includegraphics[width=0.48\textwidth]{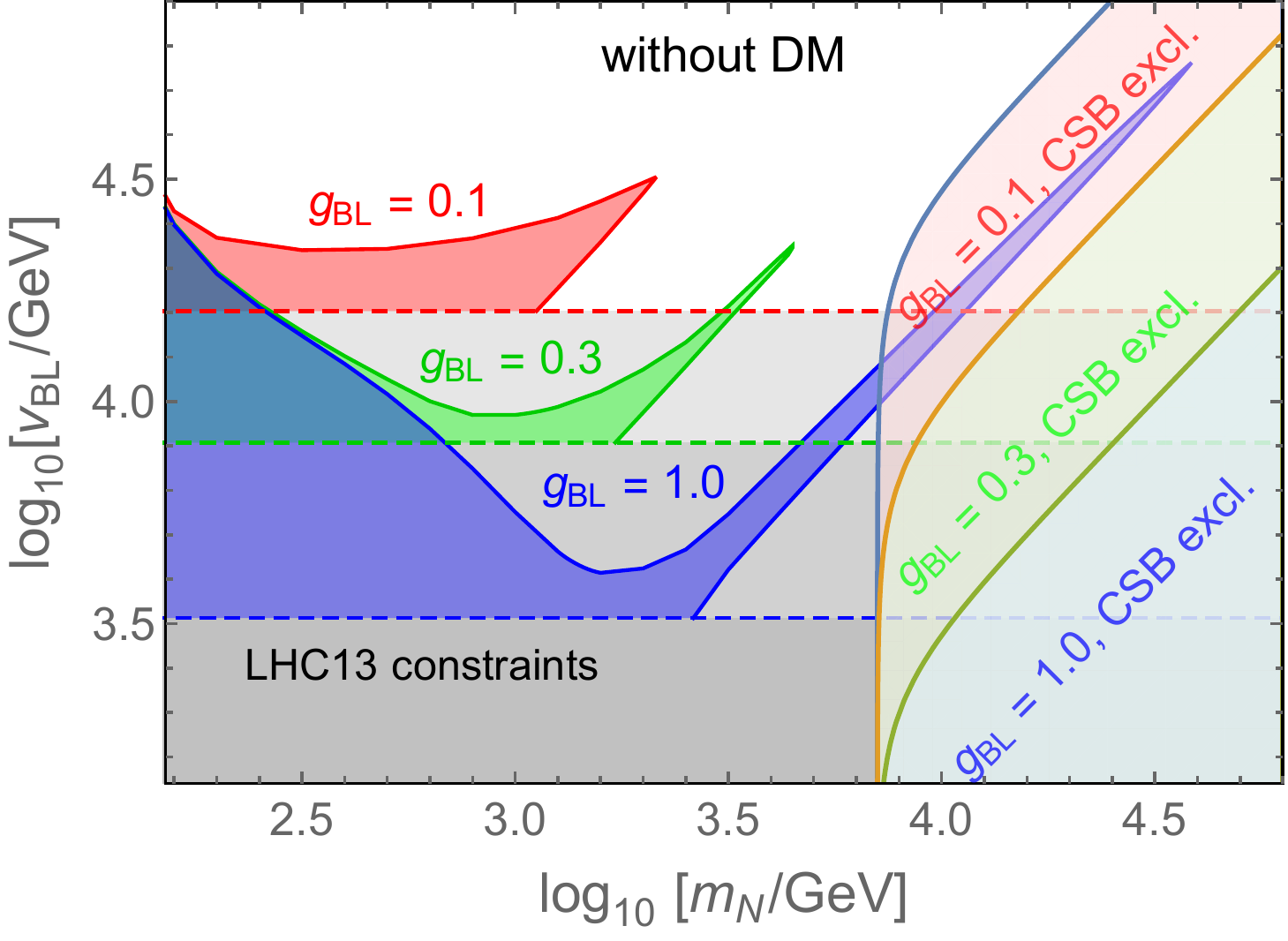}
  \includegraphics[width=0.48\textwidth]{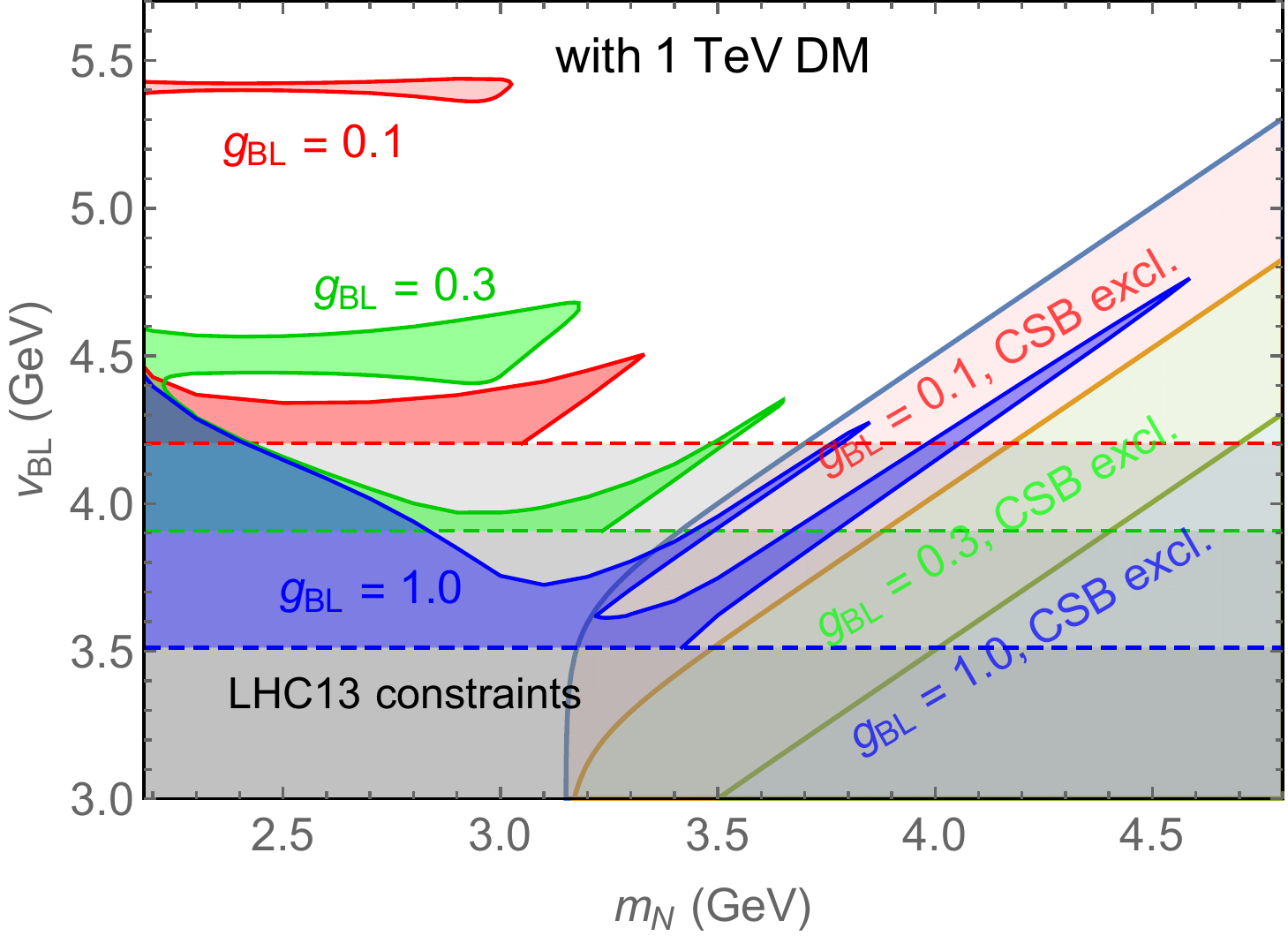}
  \caption{Parameter space for leptogenesis for the case without DM (left) and with a 1 TeV DM. The gray regions are excluded by the dilepton limits on the $Z'$ mass (cf. Fig.~\ref{fig:Zprime} and Table~\ref{tab:limits}), and the red, green and blue shaded regions are falsified by the processes shown in Fig.~\ref{fig:diagram1}, with respectively $g_{BL} = 0.1$, $0.3$ and $1$. The lighter shaded regions are excluded by the condition $m_\phi^2>0$ in Eq.~(\ref{eq:phiEWSB}). For the case with DM, we have taken the DM mass to be 1 TeV.}
  \label{fig:leptogenesis}
\end{figure}

In the case without DM, the dilution effect depends also on the RHN mass $m_N$, the $Z'$ mass $M_{Z'}$, the conformal scalar mass $m_\phi$, the quartic coupling $\lambda_\phi$, with the last three being functions of the gauge coupling $g_{BL}$ and the $B-L$ scale $v_{BL}$ in the conformal theory, as shown in Eqs.~(\ref{eqn:MZprime}), (\ref{eq:phiEWSB}) and (\ref{eqn:lambdaphi}). Therefore we choose the free parameters to be $m_N$, $g_{BL}$ and $v_{BL}$ in the conformal model, with the Yukawa coupling $Y_\phi$ determined by the RHN mass for fixed $v_{BL}$ which enters some of the diagrams in Fig.~\ref{fig:diagram1}.
%For the illustration purpose, we adopt three different benchmark values of $g_{BL} = 0.1$, $0.3$ and $1.0$ and vary $M_N$ and $v_R$ to see the implications of leptogenesis on the conformal $U(1)_{B-L}$ model. For these values of $g_{BL}$, we first interpret the solid lines in Fig.~\ref{fig:Zprime} and obtain the current dilepton constraints on the $Z'$ boson mass and the corresponding limits on $v_{BL}$, which are collected in Table~\ref{tab:limits} and
For the three benchmark values of $g_{BL} = 0.1,\, 0.3,\,1$ in Table~\ref{tab:limits}, the LHC dilepton limits on $v_{BL}$ are shown as the horizontal dashed red, green and blue lines in the left panel of Fig.~\ref{fig:leptogenesis}. As stated in Ref.~\cite{Dev:2017xry}, in the case without DM, if the RHN mass $m_N \lesssim m_\phi$ and $2m_N \lesssim m_\phi + M_{Z'}$, the dilution is dominated by the $Z'$ mediated process $NN \to f \bar{f}$, benefiting from the (almost) massless fermions in the final states and the large number of degrees of freedom. One can see the clear resonance structure in the left panel of Fig.~\ref{fig:leptogenesis}; this corresponds to the inverse decay process $N N \to Z'$ with the subsequent decay of the on-shell $Z'$ boson into SM fermions, which enhance largely the dilution effect. For heavy enough RHNs with $m_N \gtrsim m_\phi$ and/or $2m_N \gtrsim m_\phi + M_{Z'}$, the processes $NN \to \phi\phi$ and/or $NN \to Z'\phi$ are also important, which however is suppressed by the small Yukawa coupling $Y_\phi$ when $m_N \ll v_{BL}$. In the left panel of Fig.~\ref{fig:leptogenesis}, all the red, green and blue shaded regions are falsified by the extra diluting processes.

%With a (light) scalar $H_3$ coupling to the RHNs $N_i$  In the simple scenario of two quasi-degenerate RHNs, with the third one, say $N_{e,\mu}$ in the flavor space, the two RHNs have an universal coupling $f$, where for simplicity we assume the matrix $f_{ij}$ is diagonal.

When the DM mass $m_{S} \lesssim M_N$, the process $NN \to {\cal S} {\cal S}^\dagger$ would contribute to the dilution of lepton asymmetry generation,  which is mediated by the $Z'$ and $\phi$ bosons, with the Feynman diagrams shown in (f) and (g) in Fig.~\ref{fig:diagram1}. With the dashed curves in Fig.~\ref{fig:Zprime}, the dilepton limits on the $Z'$ mass and the $v_{BL}$ scale are slightly lower than the case without DM, as shown in Table~\ref{tab:limits}. The $Z'$ mediated process $NN \to Z' \to {\cal S} {\cal S}^\dagger$, however, can not compete the processes $NN \to ff$, as a result of the large degrees of freedom in the SM, unless the $B-L$ charge $n_{x}$ of DM is very large. On the other hand, the cross section $\sigma (NN \to \phi \to {\cal S} {\cal S}^\dagger)$ in the scalar portal is proportional to the trilinear scalar coupling $(\lambda_{\phi S} v_{BL})^2$, which might enhance significantly the cross section when the $v_{BL}$ scale is large. Compared to the case without DM, the new scalar portal opens the possibility of new resonance, due to the resonance relation $2E_N \simeq m_\phi$ (with $E_N$ the RHN energy) before the RHN decays. This corresponds to the extra peak structures in the right panel of Fig.~\ref{fig:leptogenesis}, where we have fixed the DM mass $m_{S} = 1$ TeV for the sake of concreteness. As in the left panel, all the red, green and blue shaded regions are falsified by the diluting processes which reduce the RHN number by two units. 
%\blue{In both the two panels of Fig.~\ref{fig:leptogenesis}, we have also shown the limits from the requirement of vacuum stability, which are shown to be the lighter shaded regions. As seen in Fig.~\ref{excl}, the regions with large RHN masses are excluded, as they have large Yukawa couplings $Y_\phi$. As demonstrated in Fig.~\ref{fig:leptogenesis}, the leptogenesis and vacuum stability constraints are largely complementary to each other. }

In short, all the gray and red, green and blue shaded regions in both of the two panels of Fig.~\ref{fig:leptogenesis} are falsified by the Feynman diagrams in Fig.~\ref{fig:diagram1}; to have a viable leptogenesis framework to generate the baryon aymmetry in the early Universe, one has to choose parameters in the unshaded regions in Fig.~\ref{fig:leptogenesis}. Roughly speaking, when the gauge coupling $g_{BL}$ (and the quartic coupling $\lambda_{\phi S}$) gets larger, the (reduced) cross sections for the dilution processes becomes larger, and the allowed parameter space shrinks significantly, depending on the RHN mass $M_N$. The bounds from the correct spontaneous symmetry breaking condition imposed by $m_\phi^2>0$ (see Eq.~(\ref{eq:phiEWSB})) are also presented in both the two panels of Fig.~\ref{fig:leptogenesis}, which exclude the large $m_N$ regions and are largely complementary to the limits from leptogenesis.

\section{WIMP DM in the conformal $U(1)_{B-L}$ model}
\label{relic}
In this section, we investigate the DM phenomenology in the conformal $U(1)_{B-L}$ model, including the relic density of DM, direct detection and collider prospects.

\subsection{Relic density and direct detection}

For the GW favored benchmark scenarios, as explored in Fig.~\ref{fig:gw}, DM annihilation at freezing-out is dominated by the process ${\cal S} {\cal S}^\dagger \to Z'Z'$, and a larger $U(1)_{B-L}$ charge $n_x$ leads to a smaller annihilation cross section, which will yield a lower value for the relic abundance of DM. 
%For  the case with of $m_{S} > M_{Z'}$, the annihilation channel $S S^\dag \to Z^\prime Z^\prime$ is active and it will dominate the DM production through freeze-out mechanism. 
The corresponding Boltzmann equation is then given by
\begin{eqnarray}
\label{eq:dYDM}
\frac{dY_S}{dx}& \ = \ &
-\frac{1}{x^2} \frac{s(m_{S})}{H(m_{S})} \Bigg[\langle \sigma v \rangle_{SS^\dag \to Z^\prime Z^\prime} \left(Y_S^2 -\frac{(Y_S^{\text{eq}})^2}{(Y_{Z^\prime}^\text{eq})^2 }Y_{Z^\prime}^2\right)\Bigg]\,,
\label{eq:dYslarge}
\end{eqnarray}
with $x\equiv m_{S}/T$ and the entropy density $s$ and Hubble parameter $H$ at the DM mass $m_S$ are resepctively
$$%\text{The ratio of entropy density $s$ and Hubble parameter $H$ at the DM mass is~~}
s(m_{S}) \ = \ \frac{2 \pi^2 }{45} g_*\, m_{S}^3, \quad 
H(m_{S}) \ = \ \frac{\pi}{\sqrt{90}} \frac{\sqrt{g_*}}{M^r_{\rm pl}} m_{S}^2 \,, $$
where $M^r_{\rm pl}= 2.44\times {10}^{18}\gev$ is the reduced Planck mass. $Y_{S,\,Z^\prime}^{\text{eq}}$ are respectively the equilibrium number densities of ${\cal S}$ and $Z^\prime$ per comoving volume. $\langle \sigma v \rangle$ in Eq.~(\ref{eq:dYDM}) \blue{is} the thermal averaged cross section and its expression is given in Appendix~\ref{sec:DMann}. After freezing-out, the total relic abundance of DM at the present epoch is obtained through the following equation~\cite{Edsjo:1997bg}
\begin{eqnarray}
\Omega_{DM} h^2 \ = \
2.755\times 10^8 \left(\frac{m_{S}}{\rm GeV}\right) Y_S(T_0) \,,
\end{eqnarray}
with $Y_S(T_0)$ obtained from numerical solutions of the coupled Boltzmann equations given in Eq. (\ref{eq:dYslarge}).

Regarding the direct detection of DM, the spin-independent (SI) process is dominated by the $Z'$ mediated scattering of DM off the nucleon ${\cal N}$, with the cross section
\begin{eqnarray}
\sigma^{\rm SI} \ \sim \
\frac{g_{BL}^4n_x^2 m_{\cal N}^2}{M_{Z'}^4} \,.
\end{eqnarray}
A large $n_x$ here can results in a large cross section, which is excluded by LUX~\cite{Akerib:2016vxi}, PandaX-II~\cite{Tan:2016zwf, Cui:2017nnn} and Xenon1T~\cite{Aprile:2017iyp}. Therefore, we do not expect the DM scalar ${\cal S}$ under study will saturate all the observed DM relic abundance.

\subsection{Collider signatures}
\label{colsig}

The DM scalar ${\cal S}$ can be produced at high-energy colliders in the scalar portal  or the gauge portal. In the scalar portal, ${\cal S}$ and ${\cal S}^\dagger$ can be pair produced through both the SM Higgs $h$ and the scalar $\phi$, assisted by the $h - \phi$ mixing which is induced by the $\lambda_P$ term in the potential (\ref{potentialcoupled4}). In particular, the most important production channel is from the gluon-fusion production of SM Higgs $h$ or $\phi$, associated with a gluon jet emitted from the initial partons, i.e.
\begin{eqnarray}
gg \to g(h/\phi) \to g {\cal S} {\cal S}^\dagger \,.
\end{eqnarray}
The DM particles ${\cal S}$ and ${\cal S}^\dagger$ leaves the detectors without leaving any signal or track, and we have a high-energy jet with large missing transverse energy at colliders. However, the production cross section is suppressed by the effective loop-level couplings of $h$ and $\phi$ to gluons, and the LHC monojet data can not set limits on the DM sector in the $U(1)_{B-L}$ model~\cite{Khachatryan:2014rra, Aad:2015zva, Aaboud:2017phn, Sirunyan:2017jix}. In the gauge portal, the most efficient way to produce DM ${\cal S}$ is from the on-shell $Z'$ decay in the process
\begin{eqnarray}
q \bar{q} \to g Z',\;\; Z' \to {\cal S} {\cal S}^\dagger \,.
\end{eqnarray}
In light of the current stringent limits on the $Z'$ boson mass~\cite{ATLAS:2016cyf,CMS:2016abv}, as shown in Fig.~\ref{fig:Zprime}, the monojet searches at LHC are too weak to set any limit on the DM sector~\cite{Khachatryan:2014rra, Aad:2015zva, Aaboud:2017phn, Sirunyan:2017jix}.

\section{Conclusion}
\label{conc}

In this paper we introduce a hidden scalar ${\cal S}$ to the $U(1)_{B-L}$ extension of the SM with classical conformal symmetry, which affects the dynamical EWSB by dimensional transmutation through the Coleman-Weinberg mechanism. The correct spontaneous symmetry breaking of the U(1)$_{B-L}$ symmetry restricts the scales of the hidden scalar and the RHNs. For smaller gauge coupling $g_{BL}$, a lower hidden scalar mass is crucial to realize a strong first order phase transition, and produce a GW signal to be probed by LISA. The possibility to realize the resonant leptogenesis mechanism is found to be disturbed by the hidden scalar depending on the mass hierarchy between it and the RHN. For the benchmark scenarios that can produce the LISA detectable GW signal and explain the baryon asymmetry of the Universe via resonant leptogenesis, we do not expect the dark matter relic abundance to be fully saturated by the hidden scalar introduced here.
%For the supercool dark matter situation scenario, we refer to Ref.~\cite{Hambye:2018qjv}. Where, the phase transition and leptogenesis mechanism are also different.

\section*{Acknowledegments}
The work of L.B. is supported by the National Natural Science Foundation of China under grant No.11605016 and No.11647307. W.C. is supported by the China Postdoctoral Science Foundation under Grant No.2019TQ0329. H.G. is partially supported by the U.S. Department of Energy grant DE-SC0009956. Y.Z. is supported by the US Department of Energy under Grant No. DE-SC0017987, and  would like to thank the Center for High Energy Physics, Peking University, the Institute of Theoretical Physics, Chinese Academy of Sciences, the Tsung-Dao Lee Institute, and the Institute of High Energy Physics, Chinese Academy of Sciences for generous hospitality where the paper was partially done.

\appendix

\section{Renormalization group equations}
\label{sec:RGE}

Following Ref.~\cite{Khoze:2014xha}, the RGEs for the scalar quartic couplings in the conformal $U(1)_{B-L}$ model read
\begin{eqnarray}
\frac{d\lambda_x}{d\log\mu}=\beta_{\lambda_x}\;,
\end{eqnarray}
with
\begin{eqnarray}
16\pi^2 \beta_{\lambda_H}& \ = \ &
-6 y_t^4+24\lambda_H^2+ \lambda_{P}^2 +\, \lambda^2_{H S} + \lambda_H \left(12y_t^2-\frac{9}{5}g_1^2-9g_2^2 \right) \nonumber\\
&&+\frac{27}{200}g_1^4+\frac{9}{20}g_2^2g_1^2+\frac{9}{8}g_2^4\;,\\
\label{eqn:beta2}
16\pi^2 \beta_{\lambda_\phi} &=& 20\lambda_\phi^2 +2\lambda_{P}^2 +\, \lambda^2_{\phi S} -48\lambda_\phi \, g_{BL}^2 +96 g_{BL}^4 \nonumber \\
&& -{\rm Tr}[Y_\phi Y_\phi^{\sf T} Y_\phi Y_\phi^{\sf T}]+8\lambda_\phi {\rm Tr}[Y_\phi Y_\phi^{\sf T}]\, \;,\\
16\pi^2 \beta_{\lambda_P}&=&\lambda_{P}\bigg(6y_t^2+12\lambda_H+8\lambda_\phi -4\lambda_{P}
-24g_{BL}^2 -\frac{9}{10}g_1^2  -\frac{9}{2}g_2^2
+4 {\rm Tr}[Y_\phi Y_\phi^{\sf T}] \bigg) \nonumber \\
&& -\,2 \lambda_{H S}\lambda_{\phi S}\;, \\
16\pi^2 \beta_{\lambda_s} &=& 20\lambda_S^2+\lambda_{\phi S}^2+2\lambda_{H S}^2+6 (n_x g_{BL})^4-12\lambda_s (n_x g_{BL})^2 \;,\\
16\pi^2 \beta_{\lambda_{HS}} &=& \lambda_{HS}\left(6y_t^2+12\lambda_H+ 6\lambda_S + 4 \lambda_{HS}
-\frac{9}{10}g_1^2-\frac{9}{2}g_2^2\right) -2\lambda_{P}\lambda_{\phi S} \;,\\
16\pi^2 \beta_{\lambda_{\phi S}}&=&
\lambda_{\phi S}\left(12\lambda_{\phi}+ 6\lambda_S
+4 \lambda_{\phi S} -18g_{BL}^2 \right)  -4\lambda_{P}\lambda_{H S}\;,
\end{eqnarray}
where $g_1^2 = 5 g_Y^2/3$ with $g_{2,\,Y}$ the gauge coupling for the SM gauge groups $SU(2)_L$ and $U(1)_Y$, $y_t$ is the SM top Yukawa coupling. For simplicity we have neglected all other Yukawa couplings in the SM as well as the couplings $Y_D$ which are much smaller. For the top quark Yukawa coupling $y_t$ and the $Y_\phi$ coupling for the three RHNs, the RGEs are respectively
\begin{eqnarray}
16\pi^2 \beta_{y_t}& \ = \ &
y_t\left(\frac{9}{2}y_t^2 -\frac{17}{20}g_1^2-\frac{9}{4}g_2^2-8g_3^2
 -\frac{2}{3}g_{BL}^2 \right) \,,  \label{ytBL} \\
16\pi^2 \beta_{Y_\phi}&=&Y_\phi\left(4Y_\phi Y_\phi^{\sf T} + {\rm Tr}[Y_\phi Y_\phi^{\sf T}]-6g_{BL}^2\right)
\label{ymBL}\,,
\end{eqnarray}
and the RGEs for the gauge couplings are given by
\begin{eqnarray}
&&16\pi^2 \beta{g_{BL}} \ = \ 12g_{BL}^3\label{eBL} \,, \quad
16\pi^2 \beta_{ g_3} \ = \ -7g_3^3 \,,\nonumber \\
&&16\pi^2 \beta{ g_2} \ = \ -\frac{19}{6}g_2^3  \,,\quad
16\pi^2 \beta{g_1} \ =  \ \frac{41}{10}g_1^3 \,,
\end{eqnarray}
where $g_3$ the gauge coupling for the SM gauge group $SU(3)_C$.

%In terms of these independent parameters, the couplings appearing in the Lagrangian can be expressed as
%\begin{eqnarray}
%\lambda_p&=&\frac{\sin(2\theta)(M_H^2-M_\Phi^2)}{2v v_{BL}},\\
%\lambda_H&=&\frac{M_H^2 cos^2\theta + M_\Phi^2 sin^2\theta}{3 v^2}+\frac{ v_{BL}\sin (2 \theta )(M_H^2-M_\Phi^2) }{12 v^3}\\
%\lambda_\phi&=&\frac{25 g_{BL}^4}{2\pi^2}+\frac{v  \sin(2 \theta )(M_H^2-M_\Phi^2)}{12v_{BL}^3}+\frac{M_H^2 sin^2\theta+M_\Phi^2\cos^2\theta}{3v_{BL}^2}
%\\
%m_\phi^2 &=&\frac{6g_{B -L}^4}{\pi^2}v_{BL}^2 + \frac{v\sin\theta \cos\theta  (M_H^2-M_\Phi^2)}{v_{BL}}\\
%\mu_H^2 & = & -\frac{v_{BL}\sin(2\theta) (M_H^2-M_\Phi^2)}{4v} ,
%\end{eqnarray}

\section{Reduced cross sections for leptogenesis}
\label{sec:leptogenesis_xs}

In this appendix, we list the explicit analytic formulas for the reduced cross sections for various $2\leftrightarrow 2$ scatterings involving the RHNs used in our leptogenesis calculations in Sec.~\ref{sec:leptogenesis}. All the relevant Feynman diagrams can be found in Fig.~\ref{fig:diagram1}. For the fermionic channels,
\begin{eqnarray}
\label{eqn:sigmaNNff}
\hat{\sigma} (NN \to f\bar{f}) & \ = \ &
\frac{S_f N_C^f (B_f -L_f)^2 g_{BL}^4}{96 \pi}
\frac{\sqrt{x} (x-4)^{3/2}}{|x-w|^2} \,,
\end{eqnarray}
with $x = s/m_N^2$, $w = M_{Z'}^2 / m_N^2$, and the symmetry factor $S_f=1$ for the charged fermions and $1/4$ for neutrinos. For the bosonic channels,
\begin{eqnarray}
%%%%%%%%%%%%%%%%%%%%%%%%%%%%%%%%%%%%%%%%%%%%%%%%%%%%%%%%%%%%
\label{eqn:H3H3}
\hat{\sigma} (NN \to \phi \phi) & \ = \ & \frac{Y_{\phi}^4}{32 \pi }
\left( {\cal A}_{SS}^{(1)} + {\cal A}_{SS}^{(2)} + {\cal A}_{SS}^{(3)} \right) \,, \\
%%%%%%%%%%%%%%%%%%%%%%%%%%%%%%%%%%%%%%%%%%%%%%%%%%%%%%%%%%%%
\hat{\sigma} (NN \to Z' \phi) &=& \frac{g_{BL}^2}{64 \pi \, w^2}
\left( {\cal A}_{VS}^{(1)} + {\cal A}_{VS}^{(2)} + {\cal A}_{VS}^{(3)} \right) \,,
\end{eqnarray}
with  the ${\cal A}_{SS}$ and ${\cal A}_{VS}$ terms
\begin{eqnarray}
{\cal A}_{SS}^{(1)} & \ \equiv \ &
\frac{121 \beta _1 (x-4) r^2}{|x-r|^2} \,, \\
%%%%%%%%%%%%%%%%%%%%%%%%%%%%%%%%%%%%%%%%%%%%%%%%%%%%%%%%%%%%
{\cal A}_{SS}^{(2)} & \ \equiv \ &
- \frac{22 r}{x (x-r)}
\bigg[ 2 \beta _1 x -
\Big(x+2 (r-4)\Big)
\log \left( \frac{(1-\beta_1)x - 2r}{(1+\beta_1)x - 2r} \right) \bigg] \,, \\
%%%%%%%%%%%%%%%%%%%%%%%%%%%%%%%%%%%%%%%%%%%%%%%%%%%%%%%%%%%%
{\cal A}_{SS}^{(3)} & \ \equiv \ &
- \beta _1 \left(1+\frac{2 (r-4)^2}{(x-2 r)^2 - \beta _1^2 x^2}\right) %\nonumber \\
%&&
- \frac{1}{2x (x-2 r)}
\Big( x^2 -4 (r-4)x + 2 (r-4) (3 r+4) \Big) \nonumber \\
&& \qquad \qquad \qquad \qquad \qquad \qquad \qquad \qquad \times \log \left( \frac{(1-\beta_1)x - 2r}{(1+\beta_1)x - 2r} \right) \,, \\
%%%%%%%%%%%%%%%%%%%%%%%%%%%%%%%%%%%%%%%%%%%%%%%%%%%%%%%%%%%%
{\cal A}_{VS}^{(1)} & \ = \ &
\frac{\beta _{3} g_{BL}^2}{(x-w)^2}
\bigg[
4 x^3 + ((w-16) w-8 r) x^2  \nonumber \\
&& \qquad + 2(2 r^2- r (w-4) w+ w^2 (3 w+10)) x \nonumber  \\
&& \qquad  + w \left(r^2 (w-8)-2 r (w-8) w+(w-40) w^2\right)
-\frac{1}{3} \beta _3^2 w^2 x^2 \bigg] \,, \\
%%%%%%%%%%%%%%%%%%%%%%%%%%%%%%%%%%%%%%%%%%%%%%%%%%%%
{\cal A}_{VS}^{(2)} & \ = \ &
-\frac{4\sqrt2 g_{BL} Y_\phi \sqrt{w}}{x (x-w)} \bigg[
\beta _3 x \left(x^2 -(r+w)x +4 w^2\right)  \nonumber \\
&& \qquad + 2 \Big(
x^2 + (r (w-2)-w (w+2))x +  r^2-r w (w+2)-(w-9) w^2
\Big) \nonumber \\
&& \qquad \qquad \qquad \qquad \qquad \qquad \times \log \left(\frac{(1-\beta_3)x - (r+w)}{(1+\beta_3)x - (r+w)}\right) \bigg] \,, \\
%%%%%%%%%%%%%%%%%%%%%%%%%%%%%%%%%%%%%%%%%%%%%%%%%%%%
{\cal A}_{VS}^{(3)} & \ = \ &
2 Y_\phi^2 w \bigg[
\beta _3 \left(x-2 w-\frac{4 (4-r) (4-w) w}{(x-r-w)^2-\beta _3^2 x^2}\right) \nonumber \\
&& \qquad -\frac{1}{x(x-r-w)} \Big( (w-2) x^2 -2(2 r (w-1)+ (w-10) w)x  \nonumber \\
&& \qquad + r^2 (w-2)+4 rw(w-1) +w ((w-10) w-32) \Big) \nonumber \\
&& \qquad \qquad \qquad \qquad \qquad \qquad \times \log \left(\frac{(1-\beta_3)x - (r+w)}{(1+\beta_3)x - (r+w)}\right) \bigg] \,,
\end{eqnarray}
where $r = m_{\phi}^2 / m_N^2$ and
\begin{eqnarray}
\beta_1 & \ = \ & \sqrt{ ( 1 - 4x^{-1} ) ( 1 - 4r x^{-1} )} \,, \\
\beta_{3} & \ = \ & \frac{1}{x}
\sqrt{ ( 1 - 4x^{-1} ) ( x^2 + r^2 + w^2 - 2xr -2xw -2r w ) } \,.
\end{eqnarray}
At the $Z'$ resonance, i.e. $M_{Z'} \simeq 2 m_N$, the propagator $1/|x-w|$ should be modified accordingly to include the $Z'$ width. For the DM channel,
\begin{eqnarray}
\label{eqn:sigmaNNdm}
\hat{\sigma} (NN \to {\cal S} {\cal S}^\dagger) & \ = \ &
\frac{\beta_2 n_{x}^2 g_{BL}^4}{768 \pi |x-w|^2}
\left[ (3-\beta_2^2)x^2 + 48y-12x(1+y) \right] \nonumber \\
&& + \frac{\beta_2 \lambda_\phi^2 (x-4)}{16\pi |x-r|^2} ,
\end{eqnarray}
with $y = m_{S}^2 / m_N^2$, and
\begin{eqnarray}
\beta_2 & \ = \ & \sqrt{ ( 1 - 4x^{-1} ) ( 1 - 4y x^{-1} )} \,.
\end{eqnarray}

\section{DM annihilation cross section}
\label{sec:DMann}

Once the kinematical threshold $m_{S}>m_{Z^\prime}$ is open, the dominant contribution to the DM pairs annihilations channel is $S (p_1) S^\dag  (p_2) \to Z^\prime (p_3) Z^\prime (p_4)$, with $p_{1,2,3,4}$ the corresponding momenta of incoming and outgoing particles. The squared amplitude is calculated using {\tt CalCHEP}~\cite{Belyaev:2012qa} with model files prepared by {\tt FeynRules}~\cite{Alloul:2013bka},
\begin{eqnarray}
|\mathcal{M}(S S^\dag\to Z^\prime Z^\prime)|^2 &=& \frac{1}{8}\bigg[ 4\lambda_{\phi S}^2 +64g_{DM}^4+ \frac{\lambda_{\phi S}^2 M_{Z^\prime}^2(16g_{BL}^4(3M_{Z^\prime}^2-s)+  \lambda_2^2 M_{Z^\prime}^2)}{g_{BL}^4(M_{H_2}^2-s)^2} + \frac{4 \lambda_2 \lambda_{\phi S}^2 M_{Z^\prime}^2}{g_{BL}^2(s - M_{H_2}^2)}
\nonumber \\
&& + \frac{4 g_{DM}^4 (6m_S^2 +3 M_{Z^\prime}^2 -2(s+t))^2}{(-m_S^2 - 2 M_{Z^\prime}^2 + s + t)^2}+ \frac{4g_{DM}^4 (-2m_S^2 + M_{Z^\prime}^2 -2t )^2}{(m_S^2-t)^2} \nonumber \\
&&- \frac{8 g_{DM}^4 (-4m_S^2 + M_{Z^\prime}^2 +s)^2}{(m_S^2-t)( m_S^2 + 2 M_{Z^\prime}^2-s - t)} + \frac{(-4m_S^2 + 2M_{Z^\prime}^2 + s -4t)}{(M_{H_2}^2-s)( m_S^2 - t)}
\nonumber \\
&&\times 8 g_{DM}^2 \lambda_{\phi S} M_{Z^\prime}^2-\frac{8 g_{DM}^2 \lambda_{\phi S} M_{Z^\prime}^2 (12m_S^2 + 6M_{Z^\prime}^2 -5s -4t)}{(M_{H_2}^2-s)( -m_S^2 -2M_{Z^\prime}^2 + s + t)}
\nonumber \\
&&+\frac{1}{(M_{H_2}^2-s)(m_S^2-t)(-m_S^2-2M_{Z^\prime}^2+s+t)}
\nonumber \\
&&\times \bigg( 8 g_{DM}^2 (g_{DM}^2 (M_{H_2}^2-s)(8 m_S^4 -16 m_S^2 t +4 M_{Z^\prime}^4-16 M_{Z^\prime}^2 t
\nonumber \\
&& -s^2 +8st +8t^2)+ 16\lambda_{\phi S} M_{Z^\prime}^2(m_S^2+2M_{Z^\prime}^2-s-t)(m_S^2 -t)) \bigg)  \bigg]\;,
\end{eqnarray}
with
\begin{eqnarray}
s&=&(p_1 + p_2)^2=(p_3 + p_4)^2\;,
\\
t&=&(p_1 - p_3)^2=(p_4 - p_2)^2\;.
\end{eqnarray}
Here, the $m_{H_2}\approx m_\phi$ for small mixing between the SM Higgs $h$ and the $B-L$ Higgs $\phi$. With the squared amplitude at hand, the cross section is given by
\begin{eqnarray}
%\sigma(S S^\dag\to h_2 h_2)&=&\frac{1}{64\pi^2}\frac{\sqrt{1-4 m_{h_2}^2/s}}{1-4 m_S^2/s}\int d\Omega |\mathcal{M}(S S^\dag\to h_2h_2)|^2\;,\\
\sigma(S S^\dag\to Z^\prime Z^\prime)&=&\frac{1}{64\pi^2}\frac{\sqrt{1-4 M_{Z^\prime}^2/s}}{1-4 m_S^2/s}\int d\Omega |\mathcal{M}(S S^\dag\to Z^\prime Z^\prime)|^2\;.
%\sigma(S S^\dag\to Z^\prime h_2)&=&\frac{1}{64\pi^2}\frac{\sqrt{(1+(m_{Z^\prime}^2-m_{h_2}^2)/s)^2-4 m_{Z^\prime}^2/s}}{\sqrt{s(1-4 m_S^2/s)}}\int d\Omega |\mathcal{M}(S S^\dag\to Z^\prime Z^\prime)|^2\;,\\
%\sigma(S S^\dag\to N_1 N_1)&=&\frac{1}{128\pi^3 }\frac{(1-4 m_{N_1}^2/s)^{3/2}}{1-4 m_S^2/s}\int d\Omega  |\mathcal{M}(S S^\dag\to Z^\prime Z^\prime)|^2\;.
\end{eqnarray}
The thermal averaged annihilation cross section $\langle{\sigma {\rm{v}}}\rangle$
are obtained in terms of annihilation cross section and the second kind modified Bessel function~\cite{Gondolo:1990dk}
\begin{eqnarray}
\left\langle \sigma {\rm v}\right\rangle
&=& \frac{1}{8 m_{S}^4 T K_2^2\left(\frac{m_{S}}{T}\right)}
\int_{4 m_{S}^2}^\infty \sigma\,(s-4 m_{S}^2)\,\sqrt{s}\,K_1
\left(\frac{\sqrt{s}}{T}\right)\,ds \,.
\label{thermal-ave}
\end{eqnarray}


\begin{thebibliography}{99}

\bibitem{Coleman:1973jx}
S.~R. Coleman and E.~J. Weinberg, ``{Radiative Corrections as the Origin of
  Spontaneous Symmetry Breaking},''
\href{http://dx.doi.org/10.1103/PhysRevD.7.1888}{{\em Phys. Rev.} {\bfseries
  D7} (1973) 1888--1910}.
%%CITATION = PHRVA,D7,1888;%%.



\bibitem{Englert:2013gz}
C.~Englert, J.~Jaeckel, V.~V. Khoze, and M.~Spannowsky, ``{Emergence of the
  Electroweak Scale through the Higgs Portal},''
  \href{http://dx.doi.org/10.1007/JHEP04(2013)060}{{\em JHEP} {\bfseries 04}
  (2013) 060},
\href{http://arxiv.org/abs/1301.4224}{{\ttfamily arXiv:1301.4224 [hep-ph]}}.
%%CITATION = ARXIV:1301.4224;%%.

\bibitem{Farzinnia:2013pga}
A.~Farzinnia, H.-J. He, and J.~Ren, ``{Natural Electroweak Symmetry Breaking
  from Scale Invariant Higgs Mechanism},''
  \href{http://dx.doi.org/10.1016/j.physletb.2013.09.060}{{\em Phys. Lett.}
  {\bfseries B727} (2013) 141--150},
\href{http://arxiv.org/abs/1308.0295}{{\ttfamily arXiv:1308.0295 [hep-ph]}}.
%%CITATION = ARXIV:1308.0295;%%.

\bibitem{Khoze:2013uia}
V.~V. Khoze, ``{Inflation and Dark Matter in the Higgs Portal of Classically
  Scale Invariant Standard Model},''
  \href{http://dx.doi.org/10.1007/JHEP11(2013)215}{{\em JHEP} {\bfseries 11}
  (2013) 215},
\href{http://arxiv.org/abs/1308.6338}{{\ttfamily arXiv:1308.6338 [hep-ph]}}.
%%CITATION = ARXIV:1308.6338;%%.

\bibitem{Iso:2009ss}
S.~Iso, N.~Okada, and Y.~Orikasa, ``{Classically conformal $B^-$ L extended
  Standard Model},''
  \href{http://dx.doi.org/10.1016/j.physletb.2009.04.046}{{\em Phys. Lett.}
  {\bfseries B676} (2009) 81--87},
\href{http://arxiv.org/abs/0902.4050}{{\ttfamily arXiv:0902.4050 [hep-ph]}}.
%%CITATION = ARXIV:0902.4050;%%.

\bibitem{Audley:2017drz}
{\bfseries LISA} Collaboration, H.~Audley {\em et~al.}, ``{Laser Interferometer
  Space Antenna},''
\href{http://arxiv.org/abs/1702.00786}{{\ttfamily arXiv:1702.00786
  [astro-ph.IM]}}.
%%CITATION = ARXIV:1702.00786;%%.

\bibitem{Cornish:2018dyw}
T.~Robson, N.~J. Cornish, and C.~Liug, ``{The construction and use of LISA
  sensitivity curves},'' \href{http://dx.doi.org/10.1088/1361-6382/ab1101}{{\em
  Class. Quant. Grav.} {\bfseries 36} no.~10, (2019) 105011},
\href{http://arxiv.org/abs/1803.01944}{{\ttfamily arXiv:1803.01944
  [astro-ph.HE]}}.
%%CITATION = ARXIV:1803.01944;%%.

\bibitem{Gong:2014mca}
X.~Gong {\em et~al.}, ``{Descope of the ALIA mission},''
  \href{http://dx.doi.org/10.1088/1742-6596/610/1/012011}{{\em J. Phys. Conf.
  Ser.} {\bfseries 610} no.~1, (2015) 012011},
\href{http://arxiv.org/abs/1410.7296}{{\ttfamily arXiv:1410.7296 [gr-qc]}}.
%%CITATION = ARXIV:1410.7296;%%.

\bibitem{Luo:2015ght}
{\bfseries TianQin} Collaboration, J.~Luo {\em et~al.}, ``{TianQin: a
  space-borne gravitational wave detector},''
  \href{http://dx.doi.org/10.1088/0264-9381/33/3/035010}{{\em Class. Quant.
  Grav.} {\bfseries 33} no.~3, (2016) 035010},
\href{http://arxiv.org/abs/1512.02076}{{\ttfamily arXiv:1512.02076
  [astro-ph.IM]}}.
%%CITATION = ARXIV:1512.02076;%%.

\bibitem{Corbin:2005ny}
V.~Corbin and N.~J. Cornish, ``{Detecting the cosmic gravitational wave
  background with the big bang observer},''
  \href{http://dx.doi.org/10.1088/0264-9381/23/7/014}{{\em Class. Quant. Grav.}
  {\bfseries 23} (2006) 2435--2446},
\href{http://arxiv.org/abs/gr-qc/0512039}{{\ttfamily arXiv:gr-qc/0512039
  [gr-qc]}}.
%%CITATION = GR-QC/0512039;%%.

\bibitem{Musha:2017usi}
{\bfseries DECIGO Working group} Collaboration, M.~Musha, ``{Space
  gravitational wave detector DECIGO/pre-DECIGO},''
\href{http://dx.doi.org/10.1117/12.2296050}{{\em Proc. SPIE Int. Soc. Opt.
  Eng.} {\bfseries 10562} (2017) 105623T}.
%%CITATION = PSISD,10562,105623T;%%.

\bibitem{Kudoh:2005as}
H.~Kudoh, A.~Taruya, T.~Hiramatsu, and Y.~Himemoto, ``{Detecting a
  gravitational-wave background with next-generation space interferometers},''
  \href{http://dx.doi.org/10.1103/PhysRevD.73.064006}{{\em Phys. Rev.}
  {\bfseries D73} (2006) 064006},
\href{http://arxiv.org/abs/gr-qc/0511145}{{\ttfamily arXiv:gr-qc/0511145
  [gr-qc]}}.
%%CITATION = GR-QC/0511145;%%.



\bibitem{Ellis:2019oqb}
J.~Ellis, M.~Lewicki, J.~M. No, and V.~Vaskonen, ``{Gravitational wave energy
  budget in strongly supercooled phase transitions},''
  \href{http://dx.doi.org/10.1088/1475-7516/2019/06/024}{{\em JCAP} {\bfseries
  1906} no.~06, (2019) 024},
\href{http://arxiv.org/abs/1903.09642}{{\ttfamily arXiv:1903.09642 [hep-ph]}}.
%%CITATION = ARXIV:1903.09642;%%.

\bibitem{Ellis:2018mja}
J.~Ellis, M.~Lewicki, and J.~M. No, ``{On the Maximal Strength of a First-Order
  Electroweak Phase Transition and its Gravitational Wave Signal},''
  \href{http://arxiv.org/abs/1809.08242}{{\ttfamily arXiv:1809.08242
  [hep-ph]}}.
[JCAP1904,003(2019)].
%%CITATION = ARXIV:1809.08242;%%.

\bibitem{Jinno:2016knw}
R.~Jinno and M.~Takimoto, ``{Probing a classically conformal B-L model with
  gravitational waves},''
  \href{http://dx.doi.org/10.1103/PhysRevD.95.015020}{{\em Phys. Rev.}
  {\bfseries D95} no.~1, (2017) 015020},
\href{http://arxiv.org/abs/1604.05035}{{\ttfamily arXiv:1604.05035 [hep-ph]}}.
%%CITATION = ARXIV:1604.05035;%%.

\bibitem{Iso:2017uuu}
S.~Iso, P.~D. Serpico, and K.~Shimada, ``{QCD-Electroweak First-Order Phase
  Transition in a Supercooled Universe},''
  \href{http://dx.doi.org/10.1103/PhysRevLett.119.141301}{{\em Phys. Rev.
  Lett.} {\bfseries 119} no.~14, (2017) 141301},
\href{http://arxiv.org/abs/1704.04955}{{\ttfamily arXiv:1704.04955 [hep-ph]}}.
%%CITATION = ARXIV:1704.04955;%%.

\bibitem{Chao:2017vrq}
W.~Chao, H.-K. Guo, and J.~Shu, ``{Gravitational Wave Signals of Electroweak
  Phase Transition Triggered by Dark Matter},''
  \href{http://dx.doi.org/10.1088/1475-7516/2017/09/009}{{\em JCAP} {\bfseries
  1709} no.~09, (2017) 009},
\href{http://arxiv.org/abs/1702.02698}{{\ttfamily arXiv:1702.02698 [hep-ph]}}.
%%CITATION = ARXIV:1702.02698;%%.

\bibitem{Chao:2017ilw}
W.~Chao, W.-F. Cui, H.-K. Guo, and J.~Shu, ``{Gravitational Wave Imprint of New
  Symmetry Breaking},''
\href{http://arxiv.org/abs/1707.09759}{{\ttfamily arXiv:1707.09759 [hep-ph]}}.
%%CITATION = ARXIV:1707.09759;%%.

\bibitem{Minkowski:1977sc}
P.~Minkowski, ``{$\mu \to e\gamma$ at a Rate of One Out of $10^{9}$ Muon
  Decays?},''
\href{http://dx.doi.org/10.1016/0370-2693(77)90435-X}{{\em Phys. Lett.}
  {\bfseries 67B} (1977) 421--428}.
%%CITATION = PHLTA,67B,421;%%.

\bibitem{Mohapatra:1979ia}
R.~N. Mohapatra and G.~Senjanovic, ``{Neutrino Mass and Spontaneous Parity
  Nonconservation},'' \href{http://dx.doi.org/10.1103/PhysRevLett.44.912}{{\em
  Phys. Rev. Lett.} {\bfseries 44} (1980) 912}.
[,231(1979)].
%%CITATION = PRLTA,44,912;%%.


\bibitem{Yanagida:1979as}
T.~Yanagida, ``{Horizontal gauge symmetry and masses of neutrinos},''
{\em Conf. Proc.} {\bfseries C7902131} (1979) 95--99.
%%CITATION = CONFP,C7902131,95;%%.

\bibitem{GellMann:1980vs}
M.~Gell-Mann, P.~Ramond, and R.~Slansky, ``{Complex Spinors and Unified
  Theories},'' {\em Conf. Proc.} {\bfseries C790927} (1979) 315--321,
\href{http://arxiv.org/abs/1306.4669}{{\ttfamily arXiv:1306.4669 [hep-th]}}.
%%CITATION = ARXIV:1306.4669;%%.

\bibitem{Glashow:1979nm}
S.~L. Glashow, ``{The Future of Elementary Particle Physics},''
\href{http://dx.doi.org/10.1007/978-1-4684-7197-7_15}{{\em NATO Sci. Ser. B}
  {\bfseries 61} (1980) 687}.
%%CITATION = HUTP-79-A059;%%.

\bibitem{Fukugita:1986hr}
  M.~Fukugita and T.~Yanagida,
  %``Baryogenesis Without Grand Unification,''
  Phys.\ Lett.\ B {\bf 174}, 45 (1986).
  %doi:10.1016/0370-2693(86)91126-3

\bibitem{Khoze:2013oga}
V.~V. Khoze and G.~Ro, ``{Leptogenesis and Neutrino Oscillations in the
  Classically Conformal Standard Model with the Higgs Portal},''
  \href{http://dx.doi.org/10.1007/JHEP10(2013)075}{{\em JHEP} {\bfseries 10}
  (2013) 075},
\href{http://arxiv.org/abs/1307.3764}{{\ttfamily arXiv:1307.3764 [hep-ph]}}.
%%CITATION = ARXIV:1307.3764;%%.

\bibitem{Covi:1996wh}
L.~Covi, E.~Roulet, and F.~Vissani, ``{CP violating decays in leptogenesis
  scenarios},'' \href{http://dx.doi.org/10.1016/0370-2693(96)00817-9}{{\em
  Phys. Lett.} {\bfseries B384} (1996) 169--174},
\href{http://arxiv.org/abs/hep-ph/9605319}{{\ttfamily arXiv:hep-ph/9605319
  [hep-ph]}}.
%%CITATION = HEP-PH/9605319;%%.

\bibitem{Flanz:1996fb}
M.~Flanz, E.~A. Paschos, U.~Sarkar, and J.~Weiss, ``{Baryogenesis through
  mixing of heavy Majorana neutrinos},''
  \href{http://dx.doi.org/10.1016/S0370-2693(96)01337-8,
  10.1016/S0370-2693(96)80011-6}{{\em Phys. Lett.} {\bfseries B389} (1996)
  693--699},
\href{http://arxiv.org/abs/hep-ph/9607310}{{\ttfamily arXiv:hep-ph/9607310
  [hep-ph]}}.
%%CITATION = HEP-PH/9607310;%%.

\bibitem{Pilaftsis:1997jf}
A.~Pilaftsis, ``{CP violation and baryogenesis due to heavy Majorana
  neutrinos},'' \href{http://dx.doi.org/10.1103/PhysRevD.56.5431}{{\em Phys.
  Rev.} {\bfseries D56} (1997) 5431--5451},
\href{http://arxiv.org/abs/hep-ph/9707235}{{\ttfamily arXiv:hep-ph/9707235
  [hep-ph]}}.
%%CITATION = HEP-PH/9707235;%%.


\bibitem{Pilaftsis:2003gt}
A.~Pilaftsis and T.~E.~J. Underwood, ``{Resonant leptogenesis},''
  \href{http://dx.doi.org/10.1016/j.nuclphysb.2004.05.029}{{\em Nucl. Phys.}
  {\bfseries B692} (2004) 303--345},
\href{http://arxiv.org/abs/hep-ph/0309342}{{\ttfamily arXiv:hep-ph/0309342
  [hep-ph]}}.
%%CITATION = HEP-PH/0309342;%%.

\bibitem{Blanchet:2009bu}
S.~Blanchet, Z.~Chacko, S.~S. Granor, and R.~N. Mohapatra, ``{Probing Resonant
  Leptogenesis at the LHC},''
  \href{http://dx.doi.org/10.1103/PhysRevD.82.076008}{{\em Phys. Rev.}
  {\bfseries D82} (2010) 076008},
\href{http://arxiv.org/abs/0904.2174}{{\ttfamily arXiv:0904.2174 [hep-ph]}}.
%%CITATION = ARXIV:0904.2174;%%.

\bibitem{Blanchet:2010kw}
S.~Blanchet, P.~S.~B. Dev, and R.~N. Mohapatra, ``{Leptogenesis with TeV Scale
  Inverse Seesaw in SO(10)},''
  \href{http://dx.doi.org/10.1103/PhysRevD.82.115025}{{\em Phys. Rev.}
  {\bfseries D82} (2010) 115025},
\href{http://arxiv.org/abs/1010.1471}{{\ttfamily arXiv:1010.1471 [hep-ph]}}.
%%CITATION = ARXIV:1010.1471;%%.

\bibitem{Iso:2010mv}
S.~Iso, N.~Okada, and Y.~Orikasa, ``{Resonant Leptogenesis in the Minimal B-L
  Extended Standard Model at TeV},''
  \href{http://dx.doi.org/10.1103/PhysRevD.83.093011}{{\em Phys. Rev.}
  {\bfseries D83} (2011) 093011},
\href{http://arxiv.org/abs/1011.4769}{{\ttfamily arXiv:1011.4769 [hep-ph]}}.
%%CITATION = ARXIV:1011.4769;%%.

\bibitem{Okada:2012fs}
N.~Okada, Y.~Orikasa, and T.~Yamada, ``{Minimal Flavor Violation in the Minimal
  $U(1)_{B-L}$ Model and Resonant Leptogenesis},''
  \href{http://dx.doi.org/10.1103/PhysRevD.86.076003}{{\em Phys. Rev.}
  {\bfseries D86} (2012) 076003},
\href{http://arxiv.org/abs/1207.1510}{{\ttfamily arXiv:1207.1510 [hep-ph]}}.
%%CITATION = ARXIV:1207.1510;%%.

\bibitem{Heeck:2016oda}
J.~Heeck and D.~Teresi, ``{Leptogenesis and neutral gauge bosons},''
  \href{http://dx.doi.org/10.1103/PhysRevD.94.095024}{{\em Phys. Rev.}
  {\bfseries D94} no.~9, (2016) 095024},
\href{http://arxiv.org/abs/1609.03594}{{\ttfamily arXiv:1609.03594 [hep-ph]}}.
%%CITATION = ARXIV:1609.03594;%%.

\bibitem{Dev:2017xry}
P.~S.~B. Dev, R.~N. Mohapatra, and Y.~Zhang, ``{Leptogenesis constraints on $B-L$ breaking Higgs boson in TeV scale seesaw models},''
  \href{http://dx.doi.org/10.1007/JHEP03(2018)122}{{\em JHEP} {\bfseries 03}
  (2018) 122},
\href{http://arxiv.org/abs/1711.07634}{{\ttfamily arXiv:1711.07634 [hep-ph]}}.
%%CITATION = ARXIV:1711.07634;%%.

\bibitem{Hambye:2018qjv}
T.~Hambye, A.~Strumia, and D.~Teresi, ``{Super-cool Dark Matter},''
  \href{http://dx.doi.org/10.1007/JHEP08(2018)188}{{\em JHEP} {\bfseries 08}
  (2018) 188},
\href{http://arxiv.org/abs/1805.01473}{{\ttfamily arXiv:1805.01473 [hep-ph]}}.
%%CITATION = ARXIV:1805.01473;%%.

\bibitem{Rodejohann:2015lca}
W.~Rodejohann and C.~E. Yaguna, ``{Scalar dark matter in the B-L model},''
  \href{http://dx.doi.org/10.1088/1475-7516/2015/12/032}{{\em JCAP} {\bfseries
  1512} no.~12, (2015) 032},
\href{http://arxiv.org/abs/1509.04036}{{\ttfamily arXiv:1509.04036 [hep-ph]}}.
%%CITATION = ARXIV:1509.04036;%%.

\bibitem{Khachatryan:2014rra}
{\bfseries CMS} Collaboration, V.~Khachatryan {\em et~al.}, ``{Search for dark
  matter, extra dimensions, and unparticles in monojet events in
  proton¨Cproton collisions at $\sqrt{s} = 8$ TeV},''
  \href{http://dx.doi.org/10.1140/epjc/s10052-015-3451-4}{{\em Eur. Phys. J.}
  {\bfseries C75} no.~5, (2015) 235},
\href{http://arxiv.org/abs/1408.3583}{{\ttfamily arXiv:1408.3583 [hep-ex]}}.
%%CITATION = ARXIV:1408.3583;%%.

\bibitem{Aad:2015zva}
{\bfseries ATLAS} Collaboration, G.~Aad {\em et~al.}, ``{Search for new
  phenomena in final states with an energetic jet and large missing transverse
  momentum in pp collisions at $\sqrt{s}=$8 TeV with the ATLAS detector},''
  \href{http://dx.doi.org/10.1140/epjc/s10052-015-3517-3,
  10.1140/epjc/s10052-015-3639-7}{{\em Eur. Phys. J.} {\bfseries C75} no.~7,
  (2015) 299}, \href{http://arxiv.org/abs/1502.01518}{{\ttfamily
  arXiv:1502.01518 [hep-ex]}}.
[Erratum: Eur. Phys. J.C75,no.9,408(2015)].
%%CITATION = ARXIV:1502.01518;%%.



\bibitem{Aaboud:2017phn}
{\bfseries ATLAS} Collaboration, M.~Aaboud {\em et~al.}, ``{Search for dark
  matter and other new phenomena in events with an energetic jet and large
  missing transverse momentum using the ATLAS detector},''
  \href{http://dx.doi.org/10.1007/JHEP01(2018)126}{{\em JHEP} {\bfseries 01}
  (2018) 126},
\href{http://arxiv.org/abs/1711.03301}{{\ttfamily arXiv:1711.03301 [hep-ex]}}.
%%CITATION = ARXIV:1711.03301;%%.

\bibitem{Sirunyan:2017jix}
{\bfseries CMS} Collaboration, A.~M. Sirunyan {\em et~al.}, ``{Search for new
  physics in final states with an energetic jet or a hadronically decaying $W$
  or $Z$ boson and transverse momentum imbalance at $\sqrt{s}=13\text{ }\text{
  }\mathrm{TeV}$},'' \href{http://dx.doi.org/10.1103/PhysRevD.97.092005}{{\em
  Phys. Rev.} {\bfseries D97} no.~9, (2018) 092005},
\href{http://arxiv.org/abs/1712.02345}{{\ttfamily arXiv:1712.02345 [hep-ex]}}.
%%CITATION = ARXIV:1712.02345;%%.

\bibitem{Akerib:2016vxi}
{\bfseries LUX} Collaboration, D.~S. Akerib {\em et~al.}, ``{Results from a
  search for dark matter in the complete LUX exposure},''
  \href{http://dx.doi.org/10.1103/PhysRevLett.118.021303}{{\em Phys. Rev.
  Lett.} {\bfseries 118} no.~2, (2017) 021303},
\href{http://arxiv.org/abs/1608.07648}{{\ttfamily arXiv:1608.07648
  [astro-ph.CO]}}.
%%CITATION = ARXIV:1608.07648;%%.

\bibitem{Tan:2016zwf}
{\bfseries PandaX-II} Collaboration, A.~Tan {\em et~al.}, ``{Dark Matter
  Results from First 98.7 Days of Data from the PandaX-II Experiment},''
  \href{http://dx.doi.org/10.1103/PhysRevLett.117.121303}{{\em Phys. Rev.
  Lett.} {\bfseries 117} no.~12, (2016) 121303},
\href{http://arxiv.org/abs/1607.07400}{{\ttfamily arXiv:1607.07400 [hep-ex]}}.
%%CITATION = ARXIV:1607.07400;%%.

\bibitem{Cui:2017nnn}
{\bfseries PandaX-II} Collaboration, X.~Cui {\em et~al.}, ``{Dark Matter
  Results From 54-Ton-Day Exposure of PandaX-II Experiment},''
  \href{http://dx.doi.org/10.1103/PhysRevLett.119.181302}{{\em Phys. Rev.
  Lett.} {\bfseries 119} no.~18, (2017) 181302},
\href{http://arxiv.org/abs/1708.06917}{{\ttfamily arXiv:1708.06917
  [astro-ph.CO]}}.
%%CITATION = ARXIV:1708.06917;%%.

\bibitem{Aprile:2017iyp}
{\bfseries XENON} Collaboration, E.~Aprile {\em et~al.}, ``{First Dark Matter
  Search Results from the XENON1T Experiment},''
  \href{http://dx.doi.org/10.1103/PhysRevLett.119.181301}{{\em Phys. Rev.
  Lett.} {\bfseries 119} no.~18, (2017) 181301},
\href{http://arxiv.org/abs/1705.06655}{{\ttfamily arXiv:1705.06655
  [astro-ph.CO]}}.
%%CITATION = ARXIV:1705.06655;%%.

\bibitem{Biswas:2016ewm}
  A.~Biswas, S.~Choubey and S.~Khan,
  %``Galactic gamma ray excess and dark matter phenomenology in a $U(1)_{B-L}$ model,''
  JHEP {\bf 1608}, 114 (2016)
  %doi:10.1007/JHEP08(2016)114
  [arXiv:1604.06566 [hep-ph]].
  %%CITATION = doi:10.1007/JHEP08(2016)114;%%
  %18 citations counted in INSPIRE as of 24 Jul 2019


\bibitem{Patra:2015bga}
S.~Patra, F.~S. Queiroz, and W.~Rodejohann, ``{Stringent Dilepton Bounds on
  Left-Right Models using LHC data},''
  \href{http://dx.doi.org/10.1016/j.physletb.2015.11.009}{{\em Phys. Lett.}
  {\bfseries B752} (2016) 186--190},
\href{http://arxiv.org/abs/1506.03456}{{\ttfamily arXiv:1506.03456 [hep-ph]}}.
%%CITATION = ARXIV:1506.03456;%%.

\bibitem{Lindner:2016lpp}
M.~Lindner, F.~S. Queiroz, and W.~Rodejohann, ``{Dilepton bounds on
  left¨Cright symmetry at the LHC run II and neutrinoless double beta
  decay},'' \href{http://dx.doi.org/10.1016/j.physletb.2016.08.068}{{\em Phys.
  Lett.} {\bfseries B762} (2016) 190--195},
\href{http://arxiv.org/abs/1604.07419}{{\ttfamily arXiv:1604.07419 [hep-ph]}}.
%%CITATION = ARXIV:1604.07419;%%.

\bibitem{ATLAS:2016cyf}
{\bfseries ATLAS} Collaboration, T.~A. collaboration,
``{Search for new high-mass resonances in the dilepton final state using
  proton-proton collisions at $\sqrt{s}$ = 13 TeV with the ATLAS detector},''.
%%CITATION = ATLAS-CONF-2016-045;%%.

\bibitem{CMS:2016abv}
{\bfseries CMS} Collaboration, C.~Collaboration,
``{Search for a high-mass resonance decaying into a dilepton final state in 13
  fb$^{-1}$ of pp collisions at $\sqrt{s}=13~\mathrm{TeV}$},''.
%%CITATION = CMS-PAS-EXO-16-031;%%.




\bibitem{Dev:2016xcp}
P.~S. Bhupal~Dev, R.~N. Mohapatra, and Y.~Zhang, ``{Naturally stable
  right-handed neutrino dark matter},''
  \href{http://dx.doi.org/10.1007/JHEP11(2016)077}{{\em JHEP} {\bfseries 11}
  (2016) 077},
\href{http://arxiv.org/abs/1608.06266}{{\ttfamily arXiv:1608.06266 [hep-ph]}}.
%%CITATION = ARXIV:1608.06266;%%.

\bibitem{Diener:2010sy}
R.~Diener, S.~Godfrey, and T.~A.~W. Martin, ``{Unravelling an Extra Neutral
  Gauge Boson at the LHC using Third Generation Fermions},''
  \href{http://dx.doi.org/10.1103/PhysRevD.83.115008}{{\em Phys. Rev.}
  {\bfseries D83} (2011) 115008},
\href{http://arxiv.org/abs/1006.2845}{{\ttfamily arXiv:1006.2845 [hep-ph]}}.
%%CITATION = ARXIV:1006.2845;%%.

\bibitem{Godfrey:2013eta}
S.~Godfrey and T.~Martin, ``{Z' Discovery Reach at Future Hadron Colliders: A
  Snowmass White Paper},'' in {\em {Proceedings, 2013 Community Summer Study on
  the Future of U.S. Particle Physics: Snowmass on the Mississippi (CSS2013):
  Minneapolis, MN, USA, July 29-August 6, 2013}}.
\newblock 2013.
\newblock \href{http://arxiv.org/abs/1309.1688}{{\ttfamily arXiv:1309.1688
  [hep-ph]}}.
\newblock
\url{http://www.slac.stanford.edu/econf/C1307292/docs/submittedArxivFiles/1309.1688.pdf}.
%\newblock
%%CITATION = ARXIV:1309.1688;%%.

\bibitem{Rizzo:2014xma}
T.~G. Rizzo, ``{Exploring new gauge bosons at a 100 TeV collider},''
  \href{http://dx.doi.org/10.1103/PhysRevD.89.095022}{{\em Phys. Rev.}
  {\bfseries D89} no.~9, (2014) 095022},
\href{http://arxiv.org/abs/1403.5465}{{\ttfamily arXiv:1403.5465 [hep-ph]}}.
%%CITATION = ARXIV:1403.5465;%%.

\bibitem{Bernon:2017jgv}
J.~Bernon, L.~Bian, and Y.~Jiang, ``{A new insight into the phase transition in
  the early Universe with two Higgs doublets},''
  \href{http://dx.doi.org/10.1007/JHEP05(2018)151}{{\em JHEP} {\bfseries 05}
  (2018) 151},
\href{http://arxiv.org/abs/1712.08430}{{\ttfamily arXiv:1712.08430 [hep-ph]}}.
%%CITATION = ARXIV:1712.08430;%%.

\bibitem{Affleck:1980ac}
I.~Affleck, ``{Quantum Statistical Metastability},''
\href{http://dx.doi.org/10.1103/PhysRevLett.46.388}{{\em Phys. Rev. Lett.}
  {\bfseries 46} (1981) 388}.
%%CITATION = PRLTA,46,388;%%.

\bibitem{Linde:1981zj}
A.~D. Linde, ``{Decay of the False Vacuum at Finite Temperature},''
  \href{http://dx.doi.org/10.1016/0550-3213(83)90293-6,
  10.1016/0550-3213(83)90072-X}{{\em Nucl. Phys.} {\bfseries B216} (1983) 421}.
[Erratum: Nucl. Phys.B223,544(1983)].
%%CITATION = NUPHA,B216,421;%%.

\bibitem{Linde:1980tt}
A.~D. Linde, ``{Fate of the False Vacuum at Finite Temperature: Theory and
  Applications},''
\href{http://dx.doi.org/10.1016/0370-2693(81)90281-1}{{\em Phys. Lett.}
  {\bfseries 100B} (1981) 37--40}.
%%CITATION = PHLTA,100B,37;%%.



\bibitem{Caprini:2015zlo}
C.~Caprini {\em et~al.}, ``{Science with the space-based interferometer eLISA.
  II: Gravitational waves from cosmological phase transitions},''
  \href{http://dx.doi.org/10.1088/1475-7516/2016/04/001}{{\em JCAP} {\bfseries
  1604} no.~04, (2016) 001},
\href{http://arxiv.org/abs/1512.06239}{{\ttfamily arXiv:1512.06239
  [astro-ph.CO]}}.
%%CITATION = ARXIV:1512.06239;%%.

\bibitem{Cai:2017cbj}
R.-G. Cai, Z.~Cao, Z.-K. Guo, S.-J. Wang, and T.~Yang, ``{The
  Gravitational-Wave Physics},''
  \href{http://dx.doi.org/10.1093/nsr/nwx029}{{\em Natl. Sci. Rev.} {\bfseries
  4} no.~5, (2017) 687--706},
\href{http://arxiv.org/abs/1703.00187}{{\ttfamily arXiv:1703.00187 [gr-qc]}}.
%%CITATION = ARXIV:1703.00187;%%.

\bibitem{Weir:2017wfa}
D.~J. Weir, ``{Gravitational waves from a first order electroweak phase
  transition: a brief review},''
  \href{http://dx.doi.org/10.1098/rsta.2017.0126}{{\em Phil. Trans. Roy. Soc.
  Lond.} {\bfseries A376} no.~2114, (2018) 20170126},
\href{http://arxiv.org/abs/1705.01783}{{\ttfamily arXiv:1705.01783 [hep-ph]}}.
%%CITATION = ARXIV:1705.01783;%%.

\bibitem{Hindmarsh:2013xza}
M.~Hindmarsh, S.~J. Huber, K.~Rummukainen, and D.~J. Weir, ``{Gravitational
  waves from the sound of a first order phase transition},''
  \href{http://dx.doi.org/10.1103/PhysRevLett.112.041301}{{\em Phys. Rev.
  Lett.} {\bfseries 112} (2014) 041301},
\href{http://arxiv.org/abs/1304.2433}{{\ttfamily arXiv:1304.2433 [hep-ph]}}.
%%CITATION = ARXIV:1304.2433;%%.

\bibitem{Hindmarsh:2015qta}
M.~Hindmarsh, S.~J. Huber, K.~Rummukainen, and D.~J. Weir, ``{Numerical
  simulations of acoustically generated gravitational waves at a first order
  phase transition},'' \href{http://dx.doi.org/10.1103/PhysRevD.92.123009}{{\em
  Phys. Rev.} {\bfseries D92} no.~12, (2015) 123009},
\href{http://arxiv.org/abs/1504.03291}{{\ttfamily arXiv:1504.03291
  [astro-ph.CO]}}.
%%CITATION = ARXIV:1504.03291;%%.

\bibitem{Jinno:2017fby}
R.~Jinno and M.~Takimoto, ``{Gravitational waves from bubble dynamics: Beyond
  the Envelope},'' \href{http://dx.doi.org/10.1088/1475-7516/2019/01/060}{{\em
  JCAP} {\bfseries 1901} (2019) 060},
\href{http://arxiv.org/abs/1707.03111}{{\ttfamily arXiv:1707.03111 [hep-ph]}}.
%%CITATION = ARXIV:1707.03111;%%.

\bibitem{Jinno:2017ixd}
R.~Jinno, S.~Lee, H.~Seong, and M.~Takimoto, ``{Gravitational waves from
  first-order phase transitions: Towards model separation by bubble nucleation
  rate},'' \href{http://dx.doi.org/10.1088/1475-7516/2017/11/050}{{\em JCAP}
  {\bfseries 1711} (2017) 050},
\href{http://arxiv.org/abs/1708.01253}{{\ttfamily arXiv:1708.01253 [hep-ph]}}.
%%CITATION = ARXIV:1708.01253;%%.

\bibitem{Jinno:2016vai}
R.~Jinno and M.~Takimoto, ``{Gravitational waves from bubble collisions: An
  analytic derivation},''
  \href{http://dx.doi.org/10.1103/PhysRevD.95.024009}{{\em Phys. Rev.}
  {\bfseries D95} no.~2, (2017) 024009},
\href{http://arxiv.org/abs/1605.01403}{{\ttfamily arXiv:1605.01403
  [astro-ph.CO]}}.
%%CITATION = ARXIV:1605.01403;%%.

\bibitem{Cutting:2018tjt}
D.~Cutting, M.~Hindmarsh, and D.~J. Weir, ``{Gravitational waves from vacuum
  first-order phase transitions: from the envelope to the lattice},''
  \href{http://dx.doi.org/10.1103/PhysRevD.97.123513}{{\em Phys. Rev.}
  {\bfseries D97} no.~12, (2018) 123513},
\href{http://arxiv.org/abs/1802.05712}{{\ttfamily arXiv:1802.05712
  [astro-ph.CO]}}.
%%CITATION = ARXIV:1802.05712;%%.

\bibitem{Kosowsky:1991ua}
A.~Kosowsky, M.~S. Turner, and R.~Watkins, ``{Gravitational radiation from
  colliding vacuum bubbles},''
\href{http://dx.doi.org/10.1103/PhysRevD.45.4514}{{\em Phys. Rev.} {\bfseries
  D45} (1992) 4514--4535}.
%%CITATION = PHRVA,D45,4514;%%.

\bibitem{Kosowsky:1992rz}
A.~Kosowsky, M.~S. Turner, and R.~Watkins, ``{Gravitational waves from first
  order cosmological phase transitions},''
\href{http://dx.doi.org/10.1103/PhysRevLett.69.2026}{{\em Phys. Rev. Lett.}
  {\bfseries 69} (1992) 2026--2029}.
%%CITATION = PRLTA,69,2026;%%.

\bibitem{Kosowsky:1992vn}
A.~Kosowsky and M.~S. Turner, ``{Gravitational radiation from colliding vacuum
  bubbles: envelope approximation to many bubble collisions},''
  \href{http://dx.doi.org/10.1103/PhysRevD.47.4372}{{\em Phys. Rev.} {\bfseries
  D47} (1993) 4372--4391},
\href{http://arxiv.org/abs/astro-ph/9211004}{{\ttfamily arXiv:astro-ph/9211004
  [astro-ph]}}.
%%CITATION = ASTRO-PH/9211004;%%.

\bibitem{Huber:2008hg}
S.~J. Huber and T.~Konstandin, ``{Gravitational Wave Production by Collisions:
  More Bubbles},'' \href{http://dx.doi.org/10.1088/1475-7516/2008/09/022}{{\em
  JCAP} {\bfseries 0809} (2008) 022},
\href{http://arxiv.org/abs/0806.1828}{{\ttfamily arXiv:0806.1828 [hep-ph]}}.
%%CITATION = ARXIV:0806.1828;%%.

\bibitem{Caprini:2009yp}
C.~Caprini, R.~Durrer, and G.~Servant, ``{The stochastic gravitational wave
  background from turbulence and magnetic fields generated by a first-order
  phase transition},''
  \href{http://dx.doi.org/10.1088/1475-7516/2009/12/024}{{\em JCAP} {\bfseries
  0912} (2009) 024},
\href{http://arxiv.org/abs/0909.0622}{{\ttfamily arXiv:0909.0622
  [astro-ph.CO]}}.
%%CITATION = ARXIV:0909.0622;%%.

\bibitem{Binetruy:2012ze}
P.~Binetruy, A.~Bohe, C.~Caprini, and J.-F. Dufaux, ``{Cosmological Backgrounds
  of Gravitational Waves and eLISA/NGO: Phase Transitions, Cosmic Strings and
  Other Sources},'' \href{http://dx.doi.org/10.1088/1475-7516/2012/06/027}{{\em
  JCAP} {\bfseries 1206} (2012) 027},
\href{http://arxiv.org/abs/1201.0983}{{\ttfamily arXiv:1201.0983 [gr-qc]}}.
%%CITATION = ARXIV:1201.0983;%%.

\bibitem{Espinosa:2010hh}
J.~R. Espinosa, T.~Konstandin, J.~M. No, and G.~Servant, ``{Energy Budget of
  Cosmological First-order Phase Transitions},''
  \href{http://dx.doi.org/10.1088/1475-7516/2010/06/028}{{\em JCAP} {\bfseries
  1006} (2010) 028},
\href{http://arxiv.org/abs/1004.4187}{{\ttfamily arXiv:1004.4187 [hep-ph]}}.
%%CITATION = ARXIV:1004.4187;%%.

\bibitem{Alves:2018jsw}
A.~Alves, T.~Ghosh, H.-K. Guo, K.~Sinha, and D.~Vagie, ``{Collider and
  Gravitational Wave Complementarity in Exploring the Singlet Extension of the
  Standard Model},'' \href{http://dx.doi.org/10.1007/JHEP04(2019)052}{{\em
  JHEP} {\bfseries 04} (2019) 052},
\href{http://arxiv.org/abs/1812.09333}{{\ttfamily arXiv:1812.09333 [hep-ph]}}.
%%CITATION = ARXIV:1812.09333;%%.

\bibitem{Alves:2018oct}
A.~Alves, T.~Ghosh, H.-K. Guo, and K.~Sinha, ``{Resonant Di-Higgs Production at
  Gravitational Wave Benchmarks: A Collider Study using Machine Learning},''
  \href{http://dx.doi.org/10.1007/JHEP12(2018)070}{{\em JHEP} {\bfseries 12}
  (2018) 070},
\href{http://arxiv.org/abs/1808.08974}{{\ttfamily arXiv:1808.08974 [hep-ph]}}.
%%CITATION = ARXIV:1808.08974;%%.

\bibitem{Kosowsky:2001xp}
A.~Kosowsky, A.~Mack, and T.~Kahniashvili, ``{Gravitational radiation from
  cosmological turbulence},''
  \href{http://dx.doi.org/10.1103/PhysRevD.66.024030}{{\em Phys. Rev.}
  {\bfseries D66} (2002) 024030},
\href{http://arxiv.org/abs/astro-ph/0111483}{{\ttfamily arXiv:astro-ph/0111483
  [astro-ph]}}.
%%CITATION = ASTRO-PH/0111483;%%.

\bibitem{Gogoberidze:2007an}
G.~Gogoberidze, T.~Kahniashvili, and A.~Kosowsky, ``{The Spectrum of
  Gravitational Radiation from Primordial Turbulence},''
  \href{http://dx.doi.org/10.1103/PhysRevD.76.083002}{{\em Phys. Rev.}
  {\bfseries D76} (2007) 083002},
\href{http://arxiv.org/abs/0705.1733}{{\ttfamily arXiv:0705.1733 [astro-ph]}}.
%%CITATION = ARXIV:0705.1733;%%.

\bibitem{Niksa:2018ofa}
P.~Niksa, M.~Schlederer, and G.~Sigl, ``{Gravitational Waves produced by
  Compressible MHD Turbulence from Cosmological Phase Transitions},''
  \href{http://dx.doi.org/10.1088/1361-6382/aac89c}{{\em Class. Quant. Grav.}
  {\bfseries 35} no.~14, (2018) 144001},
\href{http://arxiv.org/abs/1803.02271}{{\ttfamily arXiv:1803.02271
  [astro-ph.CO]}}.
%%CITATION = ARXIV:1803.02271;%%.

\bibitem{Pol:2019yex}
A.~R. Pol, S.~Mandal, A.~Brandenburg, T.~Kahniashvili, and A.~Kosowsky,
  ``{Numerical Simulations of Gravitational Waves from Early-Universe
  Turbulence},''
\href{http://arxiv.org/abs/1903.08585}{{\ttfamily arXiv:1903.08585
  [astro-ph.CO]}}.
%%CITATION = ARXIV:1903.08585;%%.

\bibitem{Brandenburg:2017neh}
A.~Brandenburg, T.~Kahniashvili, S.~Mandal, A.~R. Pol, A.~G. Tevzadze, and
  T.~Vachaspati, ``{Evolution of hydromagnetic turbulence from the electroweak
  phase transition},'' \href{http://dx.doi.org/10.1103/PhysRevD.96.123528}{{\em
  Phys. Rev.} {\bfseries D96} no.~12, (2017) 123528},
\href{http://arxiv.org/abs/1711.03804}{{\ttfamily arXiv:1711.03804
  [astro-ph.CO]}}.
%%CITATION = ARXIV:1711.03804;%%.

\bibitem{Davidson:2002qv}
S.~Davidson and A.~Ibarra, ``{A Lower bound on the right-handed neutrino mass
  from leptogenesis},''
  \href{http://dx.doi.org/10.1016/S0370-2693(02)01735-5}{{\em Phys. Lett.}
  {\bfseries B535} (2002) 25--32},
\href{http://arxiv.org/abs/hep-ph/0202239}{{\ttfamily arXiv:hep-ph/0202239
  [hep-ph]}}.
%%CITATION = HEP-PH/0202239;%%.

\bibitem{Giudice:2003jh}
G.~F. Giudice, A.~Notari, M.~Raidal, A.~Riotto, and A.~Strumia, ``{Towards a
  complete theory of thermal leptogenesis in the SM and MSSM},''
  \href{http://dx.doi.org/10.1016/j.nuclphysb.2004.02.019}{{\em Nucl. Phys.}
  {\bfseries B685} (2004) 89--149},
\href{http://arxiv.org/abs/hep-ph/0310123}{{\ttfamily arXiv:hep-ph/0310123
  [hep-ph]}}.
%%CITATION = HEP-PH/0310123;%%.

\bibitem{Dev:2014laa}
P.~S. Bhupal~Dev, P.~Millington, A.~Pilaftsis, and D.~Teresi, ``{Flavour
  Covariant Transport Equations: an Application to Resonant Leptogenesis},''
  \href{http://dx.doi.org/10.1016/j.nuclphysb.2014.06.020}{{\em Nucl. Phys.}
  {\bfseries B886} (2014) 569--664},
\href{http://arxiv.org/abs/1404.1003}{{\ttfamily arXiv:1404.1003 [hep-ph]}}.
%%CITATION = ARXIV:1404.1003;%%.

\bibitem{DOnofrio:2014rug}
M.~D'Onofrio, K.~Rummukainen, and A.~Tranberg, ``{Sphaleron Rate in the Minimal
  Standard Model},''
  \href{http://dx.doi.org/10.1103/PhysRevLett.113.141602}{{\em Phys. Rev.
  Lett.} {\bfseries 113} no.~14, (2014) 141602},
\href{http://arxiv.org/abs/1404.3565}{{\ttfamily arXiv:1404.3565 [hep-ph]}}.
%%CITATION = ARXIV:1404.3565;%%.

\bibitem{Dev:2017wwc}
B.~Dev, M.~Garny, J.~Klaric, P.~Millington, and D.~Teresi, ``{Resonant
  enhancement in leptogenesis},''
  \href{http://dx.doi.org/10.1142/S0217751X18420034}{{\em Int. J. Mod. Phys.}
  {\bfseries A33} (2018) 1842003},
\href{http://arxiv.org/abs/1711.02863}{{\ttfamily arXiv:1711.02863 [hep-ph]}}.
%%CITATION = ARXIV:1711.02863;%%.

\bibitem{Edsjo:1997bg}
J.~Edsjo and P.~Gondolo, ``{Neutralino relic density including
  coannihilations},'' \href{http://dx.doi.org/10.1103/PhysRevD.56.1879}{{\em
  Phys. Rev.} {\bfseries D56} (1997) 1879--1894},
\href{http://arxiv.org/abs/hep-ph/9704361}{{\ttfamily arXiv:hep-ph/9704361
  [hep-ph]}}.
%%CITATION = HEP-PH/9704361;%%.


\bibitem{Khoze:2014xha}
  V.~V.~Khoze, C.~McCabe and G.~Ro,
  %``Higgs vacuum stability from the dark matter portal,''
  JHEP {\bf 1408}, 026 (2014)
  %doi:10.1007/JHEP08(2014)026
  [arXiv:1403.4953 [hep-ph]].



\bibitem{Belyaev:2012qa}
  A.~Belyaev, N.~D.~Christensen and A.~Pukhov,
  %``CalcHEP 3.4 for collider physics within and beyond the Standard Model,''
  Comput.\ Phys.\ Commun.\  {\bf 184}, 1729 (2013)
  doi:10.1016/j.cpc.2013.01.014
  [arXiv:1207.6082 [hep-ph]].
  %%CITATION = doi:10.1016/j.cpc.2013.01.014;%%
  %645 citations counted in INSPIRE as of 29 Jul 2019



  \bibitem{Alloul:2013bka}
  A.~Alloul, N.~D.~Christensen, C.~Degrande, C.~Duhr and B.~Fuks,
  %``FeynRules  2.0 - A complete toolbox for tree-level phenomenology,''
  Comput.\ Phys.\ Commun.\  {\bf 185}, 2250 (2014)
  doi:10.1016/j.cpc.2014.04.012
  [arXiv:1310.1921 [hep-ph]].
  %%CITATION = doi:10.1016/j.cpc.2014.04.012;%%
  %1165 citations counted in INSPIRE as of 29 Jul 2019

\bibitem{Gondolo:1990dk}
P.~Gondolo and G.~Gelmini, ``{Cosmic abundances of stable particles: Improved
  analysis},''
\href{http://dx.doi.org/10.1016/0550-3213(91)90438-4}{{\em Nucl. Phys.}
  {\bfseries B360} (1991) 145--179}.

\end{thebibliography}
\end{document}